\documentclass[aip,jcp,reprint,groupedaddress]{revtex4-2}

\usepackage[english]{babel}
\usepackage{amsmath, amstext, amsfonts, amssymb, mathtools}
\usepackage{BOONDOX-calo}
\usepackage{dcolumn}
\usepackage{multirow}
\usepackage{booktabs}
\usepackage{bm}
\usepackage{graphicx}
\usepackage[dvipsnames]{xcolor}
\usepackage{color}
\usepackage{braket}
\usepackage{subdepth}
\usepackage[utf8]{inputenc}
\usepackage[T1]{fontenc}
\usepackage{silence} 
\usepackage{xr-hyper}
\usepackage{hyperref}
\hypersetup{pdfencoding=unicode,
   colorlinks, 
   urlcolor =blue, 
   linkcolor=black, 
   filecolor=black, 
   citecolor=black 
  }
\inputencoding{utf8}
\usepackage[nolist]{acronym}
\usepackage[sort&compress]{natbib}
\usepackage{nicefrac} 
\usepackage{footnote}
\usepackage{placeins}
\usepackage{tabularx}
\usepackage[version=4]{mhchem}
\usepackage{subfiles}
\newcommand{\supp}{Supporting Information}%

\renewcommand{\i}{\mathrm i}
\renewcommand{\a}{\alpha}
\renewcommand{\b}{\beta}

\newcommand{\cres}[2]{\hat{a}^{#2\dagger}_{#1}}
\newcommand{\anns}[2]{\hat{a}^{#2}_{#1}}
\newcommand{\ut}{\ensuremath{U(1)^2SU(2)}}
\newcommand{\uf}{\ensuremath{U(1)^4}}
\newcommand{\Be}{BeH$_2$}
\newcommand{\Fe}{[Fe(H$_2$O)$_6$]$^{2+}$}

\newcommand{\nco}{\hat{n}_{\text{C}}}
\newcommand{\nvo}{\hat{n}_{\text{V}}}
\newcommand{\nc}{n_{\text{C}}}

\usepackage{tikz}
\usetikzlibrary{shapes.symbols,arrows.meta,decorations.pathreplacing}

\definecolor{sfTEpanel}{HTML}{FFF2CC}
\definecolor{sfTEarrow}{HTML}{6C8EBF}

\newlength{\sfTEunit}
\newcommand{\spinFreeTwoETerms}[1][0.018cm]{%
  \setlength{\sfTEunit}{#1}%
  \begin{tikzpicture}[x=\sfTEunit, y=-\sfTEunit,
      elec/.style={cloud, cloud puffs=9, cloud puff arc=120, aspect=1,
        inner sep=0pt, minimum width=24\sfTEunit, minimum height=24\sfTEunit},
      occ/.style={elec, draw=none, fill=black},
      hole/.style={elec, draw=black, line width=0.4pt, fill=white},
      exc/.style={sfTEarrow, line width=5\sfTEunit,
        -{Triangle[length=16*\the\sfTEunit, width=16*\the\sfTEunit]}},
      lvl/.style={line width=0.5pt},
      lbl/.style={font=\small, inner sep=1pt},
      brc/.style={decorate, line width=0.6pt,
        decoration={brace, amplitude=7*\the\sfTEunit}}]
    \draw[brc] (44.0,-4) -- (173.8,-4)
      node[midway, above=8*\the\sfTEunit, font=\small] {\uf\unskip};
    \draw[brc] (44.0,-46) -- (329.7,-46)
      node[midway, above=8*\the\sfTEunit, font=\small] {\ut\unskip};
    \foreach \xo in {42.6, 198.5, 354.4}{
      \fill[sfTEpanel!50, rounded corners=12\sfTEunit]
        (\xo,1.5) rectangle ++(132.6,240.4);
      \foreach \yl in {16.0, 81.6, 147.0, 212.6}
        \draw[lvl] (\xo+5.8,\yl) -- ++(118,0);
    }
    \node[lbl, anchor=north west] at (44.5,16.0)  {$j$};
    \node[lbl, anchor=north west] at (44.5,81.6)  {$i$};
    \node[lbl, anchor=south west] at (44.5,147.0) {$b$};
    \node[lbl, anchor=south west] at (44.5,212.6) {$a$};
    \node[rotate=90, font=\sffamily\bfseries] at (32,180.3) {Core};
    \node[rotate=90, font=\sffamily\bfseries] at (32,53.9)  {Valence};
    \draw[exc] (77.5,199.6) -- (77.5,164.7);
    \draw[exc] (137.3,68.5) -- (137.3,33.5);
    \node[occ]  at (77.5,212.7) {};
    \node[hole] at (77.5,147.2) {};
    \node[occ]  at (137.3,81.6) {};
    \node[hole] at (137.9,16.0) {};
    \draw[exc] (233.5,198.0) -- (233.5,100.4);
    \draw[exc] (294.7,32.0) -- (294.7,131.1);
    \node[occ]  at (233.5,212.7) {};
    \node[hole] at (233.5,81.6) {};
    \node[occ]  at (293.2,16.0) {};
    \node[hole] at (293.9,147.2) {};
    \draw[exc] (387.0,199.5) -- (387.0,99.0);
    \draw[exc] (444.7,134.0) -- (444.7,33.5);
    \node[occ]  at (387.9,212.7) {};
    \node[hole] at (387.2,81.6) {};
    \node[occ]  at (444.7,147.2) {};
    \node[hole] at (444.7,16.0) {};
    \node[font=\normalsize] at (108.9,266) {$\braket{bj|ai}\,E_{ai}^{bj}$};
    \node[font=\normalsize] at (264.8,266) {$\braket{ib|aj}\,E_{aj}^{ib}$};
    \node[font=\normalsize] at (420.7,266) {$\braket{ij|ab}\,E_{ab}^{ij}$};
  \end{tikzpicture}%
}
\usepackage{tikz}
\usetikzlibrary{calc}

\newcommand{\entDir}{Figures/main}

\newcommand{\beEntPanel}[4]{%
  \begin{scope}[shift={(#1,#2)}]
    \node[anchor=south west, inner sep=0pt] at (0,0)
      {\includegraphics[width=\entWidth]{\entDir/#3}};
    \node[font=\small, anchor=south] at (0.5,1.07) {#4};
    \foreach \ang/\rad/\txt in {
        115.8/0.505/{$\sigma_\mathrm{g}^*$},
         64.3/0.495/{$1s$},
         12.9/0.485/{$\sigma_\mathrm{g}$},
        167.1/0.475/{$\sigma_\mathrm{u}^*$},
        -38.6/0.475/{$\sigma_\mathrm{u}$},
       -141.5/0.475/{$2p_y$},
        -90.0/0.475/{$2p_x$}}
      \node[font=\small, inner sep=1pt, anchor=\ang+180]
        at ($(0.5,0.5)+(\ang:\rad)$) {\txt};
  \end{scope}%
}

\newcommand{\feLabelsAll}{$4d_{-1}$, $2p_z$, $2p_y$, $2p_x$, $3d_{-1}$, $3d_{+1}$,
  $3d_{-2}$, $3d_{+2}$, $3d_{0}$, $4d_{0}$, $4d_{+2}$, $4d_{-2}$, $4d_{+1}$}
\newcommand{\feLabelsNoFourD}{{}, $2p_z$, $2p_y$, $2p_x$, $3d_{-1}$, $3d_{+1}$,
  $3d_{-2}$, $3d_{+2}$, $3d_{0}$, {}, {}, {}, {}}

\newcommand{\feEntPanel}[5]{%
  \begin{scope}[shift={(#1,#2)}]
    \node[anchor=south west, inner sep=0pt] at (0,0)
      {\includegraphics[width=\entWidth]{\entDir/#3}};
    \node[font=\small, anchor=south] at (0.5,1.10) {#4};
    \foreach \txt [count=\i from 0] in #5
      \pgfmathsetmacro{\ang}{103.9-\i*360/13}
      \node[font=\footnotesize, inner sep=0.8pt, anchor=\ang+180]
        at ($(0.5,0.5)+(\ang:0.48)$) {\txt};
  \end{scope}%
}

\newlength{\entWidth}
\newcommand{\Entanglement}[1][0.31]{%
  \setlength{\entWidth}{#1\columnwidth}%
  \begin{tikzpicture}[x=\entWidth, y=\entWidth]
    \beEntPanel{0}{1.60}{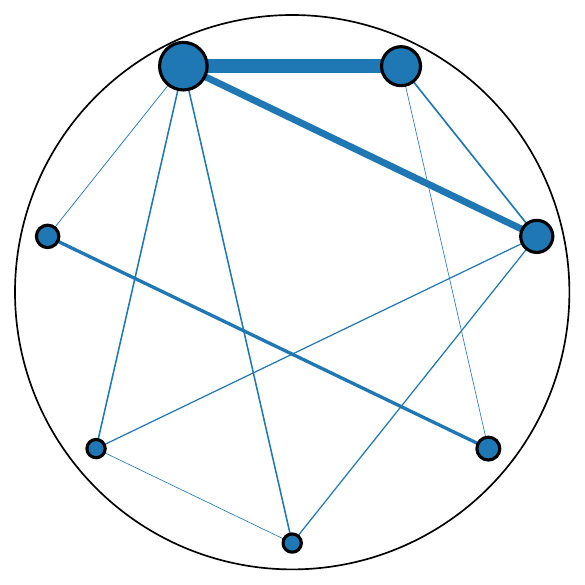}{(a) ORTHO-DMRGCI}
    \beEntPanel{1.56}{1.60}{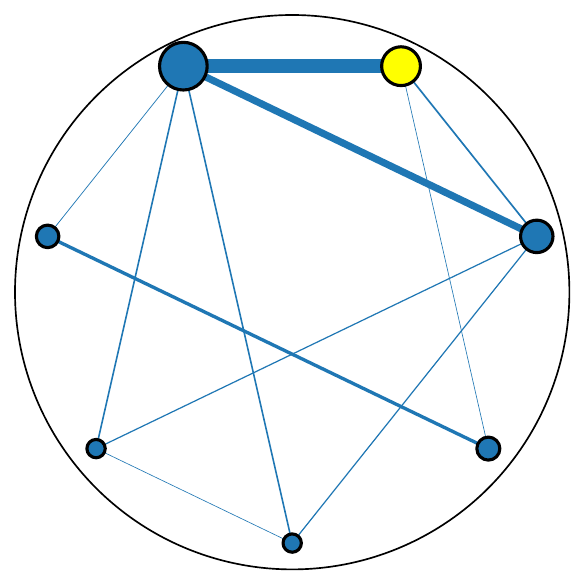}{(b) CVS-DMRGCI}
    \feEntPanel{0}{0}{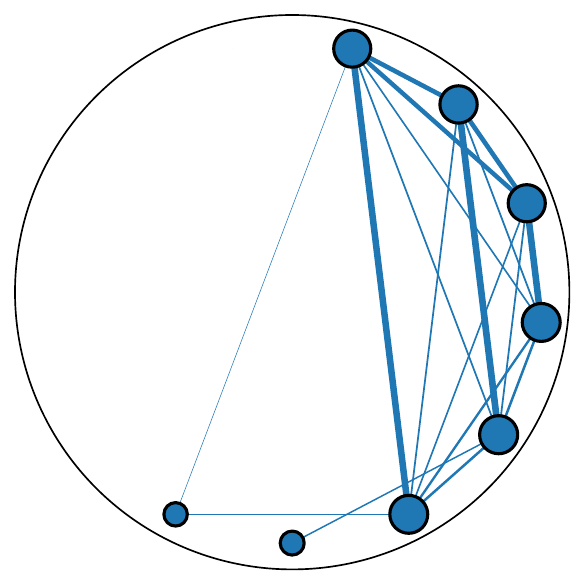}{(c) CASCI(12e,8o)}{\feLabelsNoFourD}
    \feEntPanel{1.56}{0}{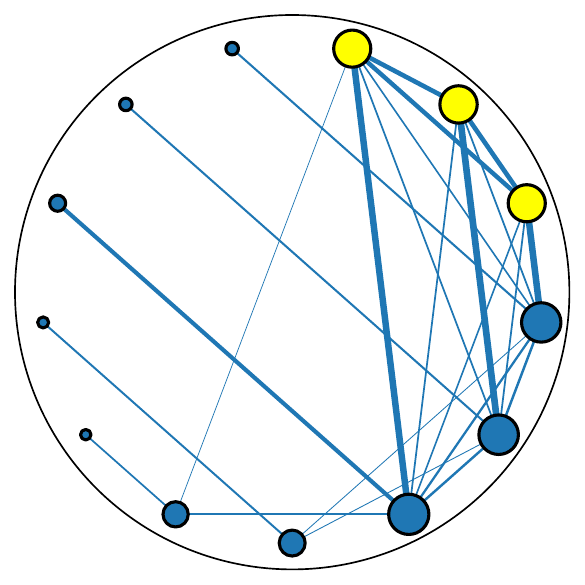}{\shortstack{(d) CVS-DMRGCI/\\HEXS-RASCI}}{\feLabelsAll}
  \end{tikzpicture}%
}

\begin{document}

\title{Density Matrix Renormalization Group Method for Core-excited Electronic States}

\author{Giacomo Bassi}
\affiliation{Chemistry Department, School of Natural Sciences, Technical University of Munich, Lichtenbergstr. 4, 85748 Garching, Germany}

\author{Mihael Erakovi\'c}
\affiliation{Division of Physical Chemistry, Ru{\dj}er Bo\v{s}kovi\'{c} Institute,
Bijeni\v{c}ka cesta 54, 10000 Zagreb, Croatia}

\author{Markus Reiher}
\affiliation{Institute of Molecular Physical Science, Department of Chemistry and Applied Biosciences, ETH Zürich, Vladimir-Prelog-Weg 1–5/10, 8093 Zürich, Switzerland}

\author{Sergey I. Bokarev}
\email[]{sergey.bokarev@tum.de}
\affiliation{Chemistry Department, School of Natural Sciences, Technical University of Munich, Lichtenbergstr. 4, 85748 Garching, Germany}
\affiliation{Institut f\"ur Physik, Universit\"at Rostock, Albert-Einstein-Str. 23-24, 18059 Rostock, Germany}

\date{\today}

\begin{abstract}
We present the formulation and implementation of the \ac{CVS} approximation within the \ac{DMRG} framework for the simulation of core-level X-ray spectra. Two symmetry implementations are developed and assessed, identifying a scheme that preserves spin symmetry and remains applicable to multiple core orbitals. Benchmark calculations for \ce{BeH2} and \ce{[Fe(H2O)6]^{2+}} demonstrate excellent agreement with conventional multi-configurational methods for excitation energies and spectral intensities, while enabling substantially larger active spaces. Using \ac{QIT} measures, we show that the \ac{CVS} constraint suppresses core-occupation fluctuations without significantly altering core--valence entanglement, providing a diagnostic for its applicability. Furthermore, we introduce a reference-free second-order perturbative correction that systematically reduces errors introduced by the \ac{CVS} approximation. Finally, a generalized multi-core formulation enables the simultaneous treatment of excitations from different core shells. These developments establish \ac{CVS}-\ac{DMRG} as an accurate and computationally efficient approach to core-level spectroscopy of strongly correlated systems.
\end{abstract}

%
%
\maketitle 
%

\begin{acronym}[RASSCF]
    \acro{ADC}{Algebraic Diagrammatic Construction}
    \acro{CAS}{Complete Active Space}
    \acro{CI}{Configuration Interaction}
    \acro{CSF}{Configuration state function}
    \acro{CVS}{Core-Valence Separation}
    \acro{DFT}{Density Functional Theory}
    \acro{DMRG}{Density Matrix Renormalization Group}
    \acro{FEAST}{}
    \acro{GAS}{Generalized Active Space}
    \acro{HEXS}{Highly Excited States}
    \acro{HF}{Hartree-Fock}
    \acro{MC}{Multi-Configurational}
    \acro{MO}{Molecular Orbital}
    \acro{MPO}{Matrix Product Operator}
    \acro{MPS}{Matrix Product State}
    \acro{MPSSI}{MPS State interaction}
    \acro{NEV}{$N$-electron valence}
    \acro{PT2}{second-order Perturbation Theory}
    \acro{QC}{Quantum Chemistry}
    \acro{QIT}{Quantum Information Theory}
    \acro{RAS}{Restricted Active Space}
    \acro{RIXS}{Resonant Inelastic X-ray Spectroscopy}
    \acro{SCF}{Self-Consistent Field}
    \acro{SD}{Slater determinant}
    \acro{SOC}{Spin-Orbit Coupling}
    \acro{WF}{Wave Function}
    \acro{XAS}{X-ray absorption spectrum}
\end{acronym}

\section{Introduction}
\label{sec:intro}

Electronic structure theory for the ground and excited states of molecules has reached a remarkable level of maturity.\cite{trygvehelgakeMolecularElectronicStructureTheory2013} 
While single-reference methods often provide an adequate description of the ground and lowest excited states, such methods are still recognized as inadequate for a broad class of problems.~\cite{Lischka_CR_2018,Park_CR_2020,Khedkar_PCCP_2021} 
Significant challenges remain when treating systems with strong electron correlation, especially in their 
excited states. 
Those effects are ubiquitous in systems exhibiting near-degenerate orbitals, such as transition metal complexes \cite{bokarevTheoreticalXraySpectroscopy2020}, lanthanide and actinide compounds \cite{aravenaPeriodicTrendsLanthanide2016}, radicals, \cite{baroneAccurateStructureThermodynamics2013} and organic molecules with extended $\pi$-systems \cite{hachmannRadicalCharacterAcenes2007} even in the ground state.\cite{kellerSelectionActiveSpaces2015}
Therefore, multi-configurational methods become essential for reliable prediction of spectra and electronic structure of such systems.

The electronic correlation problem is traditionally separated into strong (static) and weak (dynamic) correlation contributions. 
However, this conceptual division often breaks down in practice, as the two effects are inherently intertwined and difficult to disentangle. \cite{aliverti-piuriWhatCanQuantum2024} 
This observation has motivated the development of methods capable of treating large active spaces that can incorporate all essential correlation effects at the \ac{MC} \ac{SCF} level. 
Among these approaches, the \ac{DMRG} \cite{White_PRB_1993, chanDensityMatrixRenormalization2011, Schollwock_AoP_2011a}  and stochastic methods \cite{Booth_JCP_2009} have emerged as powerful tools for handling the computational cost exponentially growing  with the size of the active space associated with \ac{CAS} treatment. 

However, the situation becomes considerably more challenging when targeting highly excited states, particularly core-excited and core-ionized states. \cite{Norman_CR_2018,bokarevTheoreticalXraySpectroscopy2020,deGroot_JESRP_2021} 
For such states, the active space must be expanded to include core orbitals in addition to the valence space, thus increasing the computational cost. 
In addition, one needs to cope with the variational collapse to address the highly-lying states with a variational method without relaxing to the lowest states.
Furthermore, in processes such as resonant Auger decay from core-excited states, the manifold of final ionized states cannot typically be reproduced accurately in the wide energy range using coupled-cluster or similar single-reference methods, necessitating a truly multi-configurational treatment. \cite{Grell_PRA_2019,Grell_JCP_2020,Tenorio_JCTC_2022}  

Several strategies have been developed to address the dual challenges of large active spaces and state targeting for highly excited states. Root-homing techniques \cite{butscherModificationDavidsonsMethod1976} and energy-based shift-and-invert \cite{rakhubaCalculatingVibrationalSpectra2016}  approaches have been employed, with the harmonic Davidson method offering improvements over traditional root-homing Davidson and shift-and-invert procedures. \cite{jiaHarmonicRefinedHarmonic2015,dorandoTargetedExcitedState2007a} 
\ac{RAS} \cite{Olsen_JCP_1988b} \cite{Malmqvist_JPC_1990a} and \ac{GAS} \cite{Ma_JCP_2011a} approaches, along with others, \cite{Ivanic_JCP_2003a,Nakano_CPL_2000a} provide partial solutions by maintaining large active spaces while limiting the excitation level within orbital subspaces. This strategy simultaneously reduces computational cost and tailors the configuration expansion to describe specific classes of highly excited states. 
These methods inherently introduce a core-valence separation, which can be further enhanced by projecting out unwanted valence configurations  
in the \ac{RAS}\ac{SCF} context (HEXS-RASSCF).~\cite{delceyEfficientCalculationsLarge2019}

Despite these advances, \ac{RAS}\ac{SCF}  and \ac{GAS}\ac{SCF}  methods retain exponential scaling with the number of active orbitals. \ac{DMRG} circumvents this limitation by offering polynomial scaling, \cite{Baiardi_JCP_2020b} thereby enabling the treatment of substantially larger active spaces. Additionally, \ac{DMRG} provides a single parameter (the bond dimension) that governs both computational cost and the amount of correlation included for a given active space size. However, the \ac{RAS}/\ac{GAS} concept does not transfer naturally to \ac{DMRG} in the same straightforward manner as to the \ac{CAS} formalism, hampering its application to core-excited states.

The \ac{CVS} approximation offers an alternative approach to this problem.\cite{Cederbaum_PRA_1980b,Barth_PRA_1981b} Originally developed in the context of \ac{ADC} theory \cite{Schirmer_PRA_1982b} and later extended to other methods, \cite{Sadybekov_JCP_2017,Coriani_JCP_2015b,Norman_CR_2018b} \ac{CVS} has proven highly efficient for targeting core-excited states. \cite{Mandal_JCP_2024, Coriani_PRA_2012}
The most intuitive arguments for the separation between core and valence orbitals are, first of all, the character of spatial localization: while the core orbitals reside on a particular atom, the valence ones are delocalized over larger parts of the molecule, and secondly, the difference in energy between the two groups of \acp{MO}.\cite{Coriani_PRA_2012} 
Further, here we provide a formal analysis of the validity of the CVS approximation using \acf{QIT} quantifiers that have recently found use in the field of quantum chemistry.\cite{aliverti-piuriWhatCanQuantum2024,boguslawskiOrbitalEntanglementQuantum2015}
The \ac{CVS} approach 
naturally separates resonance states from the underlying (quasi-)continuum,~\cite{Wormit_MP_2014} and, unlike the HEXS method,~\cite{delceyEfficientCalculationsLarge2019} which works in a subspace of core configurations, \ac{CVS} avoids explicit projection onto such a subspace by instead working with an approximate Hamiltonian and a tailored initial guess, so the reachable eigenstates span precisely the desired subspace.
Thus, it is well-suited for spectroscopic applications.~\cite{rankineProgressTheoryXray2021}
Despite these advantages, \ac{CVS} has not yet been implemented within \ac{DMRG} program packages.\cite{Szenes_JPCA_2025, Block2, CheMPS2} 

On the other hand, targeting highly excited states within \ac{DMRG} presents specific algorithmic challenges that existing approaches only partially address. 
The most straightforward strategy~\cite{kellerEfficientMatrixProduct2015} consists in successive energy minimizations constrained to the orthogonal complement of all previously converged states; while reliable for low-lying excitations, it becomes impractical for core-excited states, since the projector error accumulates over the state ladder.
For instance, for singlet states computed with the $(20e, 20o)$ active space with a single core orbital in it, the first core-excited state appears beyond the lowest $1.55\cdot10^9$ of more than $5.9\cdot10^{10}$ eigenstates, 
rendering such an orthogonally-constrained variational treatment fully impractical.

Energy-specific solvers in conjunction with \ac{DMRG} offer a complementary route.~\cite{baiardiOptimizationHighlyExcited2019} 
The shift-and-invert method transforms interior eigenvalues into extremal ones by applying the spectral transformation $\Omega_\mu:= (H - \mu\mathcal{I})^{-1}$.
However, as the shift approaches an eigenvalue, the shifted operator \(H-\mu\mathcal{I}\) becomes increasingly ill-conditioned, rendering the required linear solves numerically demanding; closely spaced or degenerate target states can pose additional difficulties.~\cite{yousefsaadNumericalMethodsLarge}
The 
FEAST  algorithm~\cite{FEAST2009}, though promising for vibrational \ac{DMRG},~\cite{baiardiExcitedStateDMRGMade2022} has to date no published implementation for the electronic \ac{DMRG} case. None of these strategies provides a robust and cost-effective route to core-excited states in large \ac{DMRG} calculations, motivating the \ac{CVS}-\ac{DMRG} approach developed in this work.

In this article, we present the formulation and implementation of \ac{CVS} within the \ac{DMRG} framework. 
Specifically, we (1) develop and implement \ac{CVS}-\ac{DMRG} using two different spin symmetry treatments, (2) assess the accuracy of the \ac{CVS} approximation by comparison with {CAS-CI} and orthogonally-constrained \ac{DMRG}, labeled as ORTHO-DMRG,~\cite{kellerEfficientMatrixProduct2015} for a test system \ce{BeH2} and with RAS-CI for a transition metal complex \ce{[Fe(H2O)6]^{2+}}, (3) explore \ac{QIT} measures as diagnostics to asses the routine black-box applications,
and (4) apply second-order perturbation theory to assess \ac{CVS} accuracy and obtain  more accurate excitation energies. This work establishes \ac{CVS}-\ac{DMRG} as a practical and efficient tool for studying core-level spectroscopy in strongly correlated systems.

\section{Theory}
\label{sec:theory}

We start from a non-relativistic, spin-free molecular electronic Hamiltonian written in the restricted open-shell orbital formalism 
\begin{equation}\label{eq:SpinFreeHam}
    \hat{H} = \sum_{pq}\braket{p|\hat{h}|q}E_p^q + \frac{1}{2}\sum_{pqrs}\braket{pr|qs}E_{pr}^{qs} \, .
\end{equation}
Here, $\braket{p|\hat{h}|q}$ and $\braket{pr|qs}$ are one- and two-electron integrals and the singlet excitation operators are $E_p^q=\cres{p}{\a}\anns{q}{\a}+\cres{p}{\b}\anns{q}{\b}$ and $E_{pr}^{qs}=E_p^qE_r^s-\delta_{qr}E_p^s$.
When separating the orbitals into core (C) and valence (V) ones, the particle number operator splits into two terms
\begin{equation}\label{eq:n}
\hat{n} = \nco + \nvo = \sum_{a\in \text{C}}E_{a}^a + \sum_{i\in \text{V}} E_{i}^i \, .
\end{equation}
Below, we will use $a,b,c,d,...$ to denote core orbitals and $i,j,k,l,...$ to denote valence orbitals. 
This separation 
is arbitrary, and the number of core electrons $\nc$ is not generally conserved under the action of the molecular Hamiltonian, Eq.~\eqref{eq:SpinFreeHam}. In other words,  
any subspace particle number, e.g., $\nc$, is not a good quantum number, as electronic configurations with different $\nc$ mix. 
However, for a multi-configurational wavefunction,
this is true only for a small portion of orbitals---the active space. 
For all the other orbitals $p$ outside this space, 
the variance of $\hat{n}_p$ 
is strictly $0$, as they stay either doubly occupied or unoccupied.

The \ac{RAS}\ac{SCF} method~\cite{Malmqvist_JPC_1990a} allows defining a fixed occupation in an active orbital subspace, thus tailoring the optimization to target excited states with a certain occupation pattern. 
In particular, such an approach is useful for core-excited states.~\cite{bokarevTheoreticalXraySpectroscopy2020}
In the \ac{DMRG} context, this option, although being realizable via restricting certain sites of \ac{MPS} to, e.g., only single occupations, does not provide much computational benefit and strongly complicates the handling of data structures. 
In contrast, in this work, a concept similar, but not identical, to \ac{RAS}\ac{SCF} is implemented under the name \ac{CVS}-\ac{DMRG}.

\subsection{CVS definition}
\label{sec:cvs}
\ac{CVS} method  
partitions the active orbitals into subgroups C and V and enforces integer occupation of the core orbitals, which is different from full occupation, i.e., having one or more holes.   
Thus, the minimal condition to enforce \ac{CVS} is to search for a constrained Hamiltonian conserving ${n}_\text{C}$, i.e.,
\begin{equation}\label{eq:commHNc}
    [\hat{H}^\text{CVS},\hat{n}_\text{C}]=0 \, .
\end{equation}
To obtain $\hat H^\text{CVS}$ from the total Hamiltonian, one needs to ensure that the terms
\begin{equation}\label{eq:explicitcommHnc}
\begin{split}
&[\hat{H},\hat{n}_{\text{C}}]= 
\sum_{i}\sum_{a}h_{ia}\left(E^i_a-E^a_i
\right) + \\
&
+\sum_{a,b,c,i}\braket{ib|ac}(E^{ib}_{ac}
-E^{ac}_{ib})+\\
&
+
\sum_{a,i,j,k}
\braket{ik|aj}(E^{ik}_{aj}-E^{aj}_{ik}) +\\
&
+\sum_{a,b,i,j}\braket{ij|ab}(E^{ij}_{ab}-
E^{ab}_{ij})\\
\end{split}
\end{equation}
are zero; for detailed derivation, see \supp.
Given such a Hamiltonian and a proper wave function initial guess with the desired number of core electrons, the variational optimization of the lowest core-excited state becomes as costly as that of the ground state problem. 
One can then apply the standard methods to obtain excited states within the \ac{DMRG} formalism to address the higher core-excited states.
Thus, one efficiently avoids the problem of computing myriads of valence states 
to reach the core-excited manifold.

The accuracy of such approximation is determined by the variance of the particle number, being exact when 
\begin{equation}\label{eq:CorePopVarZero}
   \braket{\nco^2}_{\Psi}-\braket{\nco}_{\Psi}^2=0\, ,
\end{equation}
where $\braket{\dots}_\Psi$ denotes the expectation value on a wave function $\Psi$.
Thus, the orbitals naturally fulfilling Eq.~\eqref{eq:CorePopVarZero} for the full set of Hamiltonian eigenstates as closely as possible are best suited for being selected as core in the \ac{CVS} treatment.

Importantly, although the \ac{CVS} approximation implicitly restricts the accessible configuration space, it does not generally enforce a separable wavefunction with respect to the C and V subsystems. 
As will be shown in this work, core–valence entanglement persists in a CVS wavefunction $\Psi_\text{CVS}$, as quantified by a nonzero bipartite entanglement entropy,
\[
S_{C|V,\Psi_{\text{CVS}}}>0 \, .
\]
Thus, CVS does not eliminate core–valence correlation, although it somewhat restricts its contribution.

Note that the version of \ac{CVS} in Eq.~\eqref{eq:commHNc} applies constraints on the Hamiltonian and directly ensures the core electron number conservation. 
Thus, it is different from the version applied in other works, aiming at decoupling valence, single-, and double-core-hole blocks of the Hamiltonian.
In this case, only the two-electron integrals in Eq.~\eqref{eq:explicitcommHnc} are zeroed.  
By construction, such a version of CVS is projective in nature; see, e.g., Ref.~\citenum{Norman_CR_2018}. 
In this sense, the variant of \ac{CVS} used here is more general and can be introduced as an additional continuous symmetry.

\subsection{CVS as an additional DMRG symmetry}

Particle-number conservation corresponds to a $U(1)$ symmetry, which can be explicitly exploited in \ac{MPS}/\ac{MPO} representations.~\cite{McCulloch_JSM_2007}
In addition to this, the non-relativistic molecular Hamiltonian exhibits total spin conservation, referred to as $SU(2)$ symmetry ($[\hat H^\text{CVS}, \hat S^2]=0$). The \ac{CVS} approximation corresponds to the additional conservation of particles in the core and valence orbital subspaces separately. Thus, we denote the \ac{CVS} symmetry as $U(1)$ and use the already existing capabilities of the symmetry-adapted \ac{DMRG} \cite{kellerSpinadaptedMatrixProduct2016} to implement \ac{CVS}.

\subsubsection{\texorpdfstring{\uf\ symmetry}{U(1)4 symmetry}}

One way to introduce \ac{CVS} is to subdivide each of the C and V subspaces into C$_\alpha$/C$_\beta$ and V$_\alpha$/V$_\beta$.
For this, one splits the sums in Eq.~\eqref{eq:n} 
\begin{equation}\label{eq:4n}
\hat{n} = \hat n_{\text{V}\alpha} +\hat n_{\text{V}\beta}+\hat n_{\text{C}\alpha}+\hat n_{\text{C}\beta} \, .
\end{equation}
Conserving the number of spin-up and spin-down electrons within a subspace independently formally corresponds to the $U(1)_{V\alpha}\times U(1)_{V\beta}\times U(1)_{C\alpha}\times U(1)_{C\beta}$ symmetry; 
here, this approach is referred to as \uf\ symmetry. 
This corresponds to constructing \ac{MPS} with non-correlated allowed sectors characterized as $\ket{n_{\text{V}\alpha},n_{\text{V}\beta},n_{\text{C}\alpha},n_{\text{C}\beta}}$ and building an MPO that has the same block structure to avoid mixing within sectors. 
In practice, this is done by masking (zeroing) the one- and two-electron integrals in Eq.~\ref{eq:explicitcommHnc}.

However, when applied to \ac{DMRG}, the \uf approach 
has several shortcomings. 
First of all, the MPSs in the form of $\ket{n_{\text{V}\alpha},n_{\text{V}\beta},n_{\text{C}\alpha},n_{\text{C}\beta}}$ are not eigenstates of $\hat S^2$. Secondly, in the construction of the MPO, more terms than necessary are set to zero, see below. Hence, the \uf symmetry corresponds to an additional restriction to $\hat H^{\text{CVS}}$ beyond the minimal approximation to the full $\hat H$.
{Despite this, we foresee the utility of the \uf\ implementation for the cases when C and V orbital groups are spatially separated, and each of the subsystems can be represented in a localized-spin (diabatic) basis, e.g., for multi-center magnetic molecules~\cite{Wodynski_JCTC_2018, Safari_JCTC_2025} which undergo non-local electron decay.~\cite{Jahnke_CR_2020}}

\subsubsection{\texorpdfstring{\ut\ symmetry}{U(1)2SU(2) symmetry}}
\label{sec:ut}

To solve both problems of \uf, 
the \ac{CVS} Hamiltonian should conserve the four sums 
\begin{equation}
\begin{split}
    & \hat{n}_{\text{V}} = \hat n_{\text{V}\alpha} +\hat n_{\text{V}\beta} \qquad 
    \hat{n}_{\text{C}} =\hat n_{\text{C}\alpha}+\hat n_{\text{C}\beta} \\
    &
    \hat{n}_{\alpha} = \hat n_{\text{V}\alpha} +\hat n_{\text{C}\alpha} \qquad
    \hat{n}_{\beta} =\hat n_{\text{V}\beta}+\hat n_{\text{C}\beta} \, ,
\end{split}
\label{eq:2n2}
\end{equation}
instead of conserving each particle number in Eq.~\eqref{eq:4n}.
However, due to linear dependence $\hat n_{\mathrm C}+\hat n_{\mathrm V}=\hat n_\alpha+\hat n_\beta$, {one can express such a symmetry as $U(1)_{n_V}\times U(1)_{n_C} \times U(1)_{S_z}$. 
Note that there is no restriction on the spin of the C or V subspaces separately, and core-valence spin exchange remains possible (Fig.~\ref{fig:4u1vs22u1Diagram}); thus, for a spin-independent Hamiltonian, this restriction preserves the full spin-rotation symmetry, leading to a $U(1)_{n_{\mathrm C}}\times U(1)_{n_{\mathrm V}}\times SU(2)_{\mathrm{spin}}$ symmetry group; thus, the name \ut.
The \ac{MPS} allowed sectors are then defined as $\ket{n_{\text{V}},n_{\text{C}},S}$ and the MPO is again constructed, masking the terms in Eq. ~\ref{eq:explicitcommHnc}.

The difference between \ut\ and \uf\ lies in the treatment of the fourth term in Eq.~\ref{eq:explicitcommHnc} and the use of integral and excitation operator symmetries with respect to index permutations.
These differences are depicted graphically in Fig.~\ref{fig:4u1vs22u1Diagram}.
The excitation operator symmetry is also important to understand the difference between our formulation and other \ac{CVS} implementations.~\cite{Norman_CR_2018}

All three formulations, \uf, \ut, and from the literature, neglect the $\braket{ij|ab}E^{ij}_{ab}$ terms.
However, the \ut\ formulation includes the core-valence exchange terms $\braket{ib|aj}E^{ib}_{aj}$, whereas the \uf\ neglects them;
note that $\braket{ij|ab}E^{ij}_{ab}\neq\braket{ib|aj}E^{ib}_{aj}$ but $\braket{ij|ab}=\braket{ib|aj}$ due to integral and excitation operator symmetry.
\begin{figure}[t]
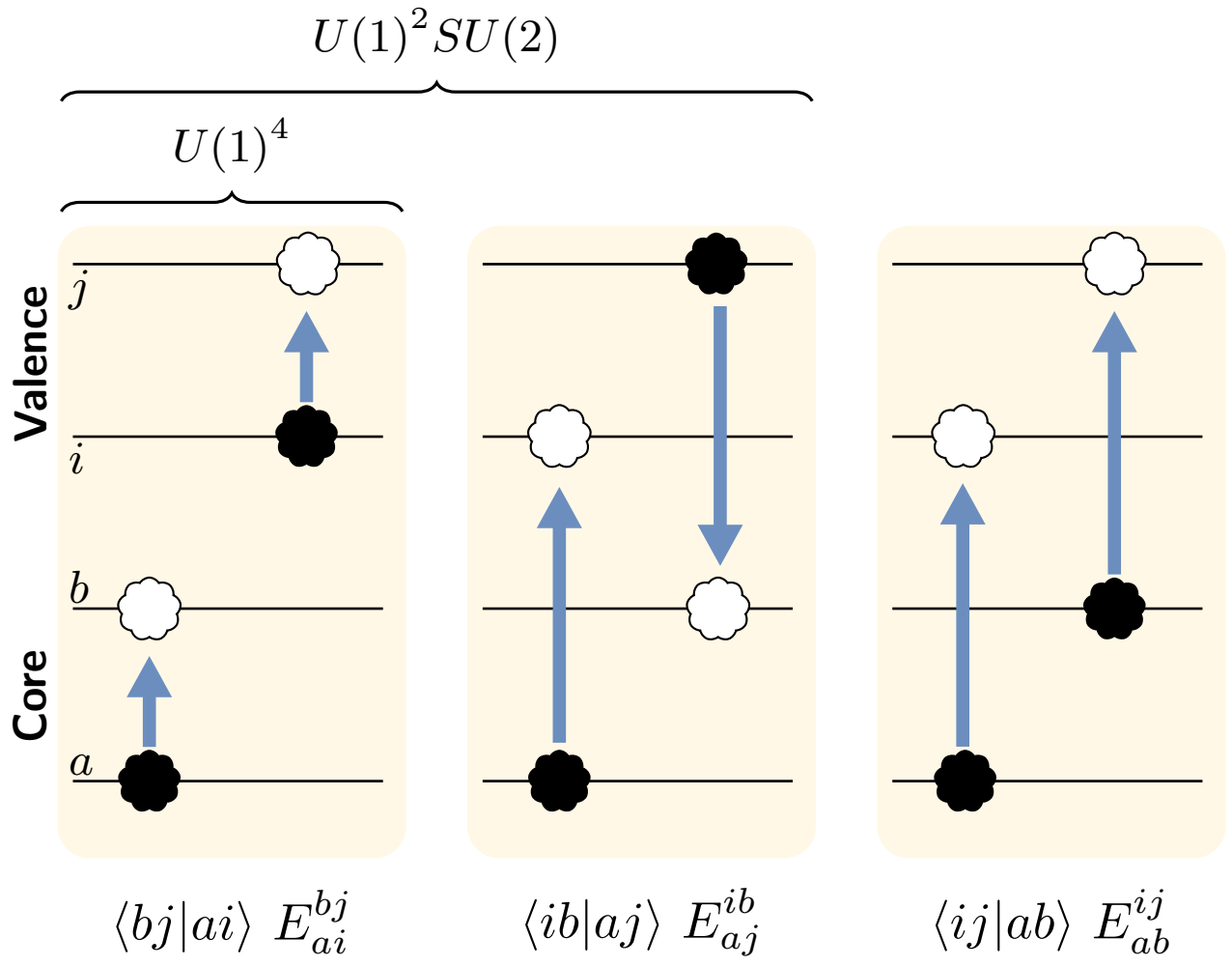

    \centering
    \spinFreeTwoETerms
    \caption{
    The graphical representation of the different terms in the \ut\ and \uf\ \ac{CVS}-Hamiltonians arising from the fourth term in Eq.~\eqref{eq:explicitcommHnc}, when accounting for index permutation symmetry of integrals and excitation operators. 
    The arrows denote the ``direction'' of the excitation operators.
    } 
    \label{fig:4u1vs22u1Diagram}
\end{figure}
Due to the fact that $\hat H^{\text{CVS}}(\uf)$ is more restrictive than $\hat H^{\text{CVS}}(\ut)$, in the cases when there are several core orbitals, the number of terms which are neglected in \uf\ increases dramatically. 
To give a perspective, for a ten-orbital valence space and a single core orbital, there are 200 terms of difference between \uf\ and \ut, while for the same valence space and three core orbitals the difference grows to 1800.
Thus, one can expect larger deviations between the two formulations for, e.g., L$_{2,3}$ or $M_{2-5}$ spectra.
In general, the \ut\ and the literature implementation~\cite{Norman_CR_2018} are equivalent, if one lifts the 8-fold index-permutation symmetry of integrals and accounts for the two-fold symmetry of the excitation operators.  

Since $\ket{n_{\text{V}},n_{\text{C}},S}$ are eigenstates of $\hat S^2$, they can be directly used as an initial guess for $SU(2)$-DMRG optimizations; on the contrary, one needs a combination of  $\ket{n_{\text{V}\alpha},n_{\text{V}\beta},n_{\text{C}\alpha},n_{\text{C}\beta}}$ to initialize a spin-pure MPS, similarly to how one mixes \acl{SD} into configuration state functions.
Because of the above advantages of \ut, and since no computational cost reduction has been observed for \uf, we further mainly discuss the \ut\ formulation,
referring to the \ut\ symmetry as \ac{CVS} symmetry if not stated otherwise. 

\subsection{Multi-core CVS-DMRG} 

The CVS approximation can be generalized 
to a multipartite active space,
where one introduces $N$ non-overlapping subspaces $\{\text{I}\}$, each with its own occupation-number operator $\hat{n}_\text{I} = \sum_{i\in\text{I}} E_i^i$.
The multi-core CVS Hamiltonian is then defined by requiring that each subspace occupation is separately conserved,
\begin{equation}\label{eq:multi_commHNc}
    [\hat{H}^\text{CVS}, \hat{n}_\text{I}] = 0 \qquad \forall\;\text{I}\, .
\end{equation}

This generalizes the bipartite \ac{CVS} condition Eq.~\eqref{eq:commHNc} to an arbitrary number of core shells, allowing one to target, for example, simultaneous $K$- and $L$-edge excitations or multi-site core-excited states in a \ac{DMRG} calculation. 
As the inclusion of multiple core subspaces can notably increase the size of the active space, the advantages offered by the \ac{DMRG} method come to the forefront.
\ac{CVS} might serve as a good approximation for describing cases containing not only core and valence orbitals but also possessing spatial locality, e.g., in cases of coupled subsystems. 

\subsection{Perturbative correction for CVS error } 
\label{subsec:PT2_theory}
To account for the error introduced by the CVS approximation, a perturbative correction is constructed by coupling CVS states with a given number of core electrons to states 
with a different number. Within the first-order perturbation theory, the corrected wavefunction acquires contributions from states $\Psi^{(n_C', n_V')}$ with $(n_C', n_V') \neq (n_C, n_V)$, mediated by the residual Hamiltonian $\hat H':=\hat{H} - \hat{H}^\text{CVS}$.
Since the Hamiltonian does not contain terms beyond two-body operators, the sum over perturbing manifolds reduces to only those with $n_C' = n_C \pm 1$ and $n_C' = n_C \pm 2$. Furthermore, all matrix elements between states belonging to the same $(n_C, n_V)$ manifold vanish exactly, since $\hat{H}'$ carries, by construction, no coupling within configuration sectors with the same number of core electrons.
Thus, the standard non-degenerate Rayleigh--Schr\"{o}dinger expansion is well-defined regardless of accidental state degeneracies within one manifold.

The second-order energy correction for a state $I$ in the $(n_C, n_V)$ manifold reads:
\begin{equation}\label{eq:PT2}
    E^{(2)}_I = \sum_{n_C' = n_C \pm 1,\, \pm 2} \sum_{J \in \mathcal{M}(n_\text{C}', n_\text{V}')}
    \frac{\left|\langle \Psi_J \mid \hat{H}' \mid \Psi_I \rangle\right|^2}{E_I^{(0)} - E_J^{(0)}}
\end{equation}
Because of relatively small values of couplings in the numerator and large energetic separation or valence, single-, double-, and triple-core-hole manifolds in the denominator, the perturbation expansion is well justified.

\section{ Methods and computational details}
\label{sec:comp_details}
All calculations were performed using the local developers' version of the \textsc{OpenMolcas} package~\cite{LiManni_JCTC_2023b} interfaced with the developers' version of \textsc{QCMaquis}~\cite{Szenes_JPCA_2025} for the \ac{DMRG} step. 
Scalar relativistic effects were accounted for via the second-order Douglas--Kroll--Hess  Hamiltonian~\cite{Douglas_AP_1974,Hess_PRA_1986} in both systems. For \Be, the ANO-RCC basis set~\cite{Widmark_TCA_1990,Roos_JPCA_2005} was employed with VTZP quality on Be and VDZ on H; for \Fe, VTZP contraction was used on Fe and O and VDZ on H.

For \Be, the (6e,7o) active space included the Be 1s orbital plus the full valence space: the combinations of Be 2s and three 2p orbitals together with the two H 1s orbitals.
For \Fe, the (6e,10o) active space, comprising the valence $3d$ orbitals and the $4d$ orbitals accounting for the double shell effect,~\cite{botchValenceCorrelation21981,Anderson_CPL_1992} i.e., $3\times n3d$, $2\times\sigma^\ast 3d$, and $5\times n4d$,
was employed as the basic valence space. 
Depending on the targeted core-excited edge, additional core orbitals were added to the active space: 
\begin{itemize}
    \item AS(8e,11o): $1s$+AS(6e,10o); for the $K$-edge.
    \item AS(12e,13o): $3\times2p$+AS(6e,10o); for the $L_{2,3}$-edge.
    \item AS(12e,13o): $3\times3p$+AS(6e,10o); for the $M_{2,3}$-edge.
\end{itemize}
{For selected cases, we have also performed calculations where the double-shell $4d$ orbitals were excluded from AS(12e,13o); these results will be denoted (12e,8o) in what follows.}

As a reference to compare with DMRG results, especially for \Fe, where ORTHO-DMRG was not feasible, additional \ac{CAS}-\ac{CI}, \ac{RAS}-\ac{CI}, and \ac{RAS}\ac{SCF} calculations were done using the same active spaces  
The core orbitals were put into the RAS1 subspace, allowing for one core-hole and employing the projection onto the core-excited manifold (HEXS).~\cite{delceyEfficientCalculationsLarge2019} Other orbitals were in the RAS2 subspace.

For DMRG-CI calculations of \Be, reference orbitals were obtained through a state-specific \ac{CAS}\ac{SCF} calculation of the singlet ground state. 
For the spectral calculations of \Fe, the orbital relaxation upon spin change and core excitation appeared essential. 
Thus, the orbitals were optimized at the CVS-DMRGSCF level for the lowest state in each spin and valence/core manifold, since the state-averaged DMRGSCF involving multiple states appeared to be very unstable. 
RASSCF calculations followed the same logic.
However, to correct for the \ac{CVS} error using perturbation theory, all the manifolds with different $n_\text{C}$, see Eq.~\eqref{eq:PT2}, were computed at the DMRG-CI level with fixed orbitals, using the lowest singlet valence state orbitals of \Fe.
In \ac{MPS}, the orbitals were arranged according to their canonical energies. 
The convergence tests with respect to the MPS bond dimension confirmed that the chosen ordering provides sufficiently accurate results.

The \ac{CVS}-\ac{DMRG} and the ORTHO-DMRG calculations were performed using $SU(2)$-symmetric \ac{MPS} with  maximum bond dimension $m=256$ for \Be\ and \Fe.
{During the sweeping, at each site, the discarded weight and the effective bond dimension were monitored. For \Be\ and the \Fe\ valence states, the maximum bond dimension was never reached, and the discarded weight stayed below the truncation threshold of $10^{-16}$. For \Fe, core-excited states, the maximum bond dimension was reached, but the comparison with  RASSCF results and an additional run with $m=1024$ confirmed a satisfactory bond dimension convergence with $m=256$; see {Table~S1 and Fig.~S1} in the \supp. The sweeping and Davidson convergence thresholds were set to default values, while the maximum number of sweeps and Davidson iterations were set so that such limits were never reached and the convergence was ruled only by the thresholds. 
The latter was found to be particularly important to be increased from the default value of 10, since poorly converged two-site problems frequently led to numerical instabilities and problems in the overall optimization.}

The first state of each manifold (core/valence and different spin) was obtained as the ground eigenstate of the $\hat H^\text{CVS}$ with the respective $n_\text{C}$. All other (core-)excited states were calculated using the ORTHO-DMRG scheme on top of it. Using the $\hat H^\text{CVS}$ on both valence and core states appeared to be essential: if a non-CVS Hamiltonian is used for valence manifolds, this leads to additional energy shifts; {see below and \supp}.
For \Be, 10 valence and 30 core singlet states were calculated separately using (HEXS)-RASSCF and CVS-DMRG.
For ORTHO-DMRG and CASSCF, 135 total singlet states, of which 105 are valence and 30 are core, were computed. 
For \Fe, 45 valence triplet, 5 valence quintet, 150 core triplet, 30 core quintet were computed for the $L$-edge at the CVS-DMRG level. 
For the $M$-edge, 30 valence and 30 core quintet states were computed.
The same states were also used in (HEXS)-RASSCF calculations for comparison.
Electronic absorption spectra and transition properties were computed using the \textsc{RASSI} and \textsc{MPSSI}~\cite{Knecht_JCTC_2016,Freitag_JCTC_2021} modules within \textsc{OpenMolcas}, with the atomic mean-field integral treatment of \ac{SOC}.~\cite{Malmqvist_CPL_2002}
To facilitate the spectroscopic calculations for large manifolds of states, we have skipped the biorthogonalization of orbitals in \textsc{MPSSI}.

The one- and two-particle reduced density matrices were computed with the \textsc{measure[ChemEntropy]} tool embedded in QCMaquis software.~\cite{Szenes_JPCA_2025} 
The CVS-PT2 energy corrections, number expectation values, and their variance were computed using \textsc{python} scripts available online.~\cite{PythonScripts}

\section{Results and discussion}
\label{sec:results}




The new implementation substantially extends the range of active-space sizes accessible to practical calculations. However, a comprehensive assessment of numerical accuracy becomes increasingly challenging for such large problems. Therefore, in this work, we establish the reliability of the approach through systematic study on smaller, computationally tractable systems.

In this section, we consider the core-excited states and X-ray absorption spectra (\acs{XAS}) of both systems---\Be\ ($K$-edge) and \Fe\ ($L_{2,3}$- and $M_{2,3}$-edges) for a series of methods.
Later in this section, the validity of the \ac{CVS} approximation from the viewpoint of \ac{QIT} and perturbation theory correction is assessed.
However, we begin by examining the numerical aspects of CVS-DMRG calculations for core-excited states, focusing on the methodological choices and convergence tests underlying the results presented below.

\subsection{Methodological Assessment of CVS-DMRG}

\paragraph{\uf\ accuracy}
{The comparison of the \uf\ and \ut\ results is shown in 
{Fig.~S2}
in \supp} for both test systems. 
In general, \uf\ gives higher energies of the core-excited manifold as a whole and, although the order of states is kept, it reshuffles the energy spacings between states and lifts degeneracies. 
This behavior is notably more pronounced when more than one orbital is in the core orbital subspace, i.e., the \Fe\ $L_{2,3}$-edge, due to neglect of more exchange terms in \ut; see Sec.~\ref{sec:ut}. 
A more detailed comparison of the \uf\ and \ut\ in terms of \ac{QIT} can be found in \supp\ {Figs.~S3 and S4}. 
Since the \uf\ formulation yields less accurate energies and distorts orbital entanglement patterns, we restrict the following discussion to the \ut\ formulation,
although we admit that \uf\ might be useful in some contexts.

\paragraph{ORTHO-CVS-DMRG}
\label{sec:DMRG_problems}
Although DMRG is a variational method, the optimization of MPS tensors constitutes a non-linear, non-convex problem, and convergence to the global energy minimum is not formally guaranteed.~\cite{Schollwock_AoP_2011a, Catarina_EPJB_2023}
Thus, even ground-state DMRG can exhibit local minima, metastability, or convergence to suboptimal states.
Excited states introduce an additional difficulty; although they also possess a well-defined global variational minimum, provided orthogonality to all lower states, these lower eigenstates are generally unknown and must themselves be approximated.
Thus, additional problems arise from root flipping, imperfect orthogonality constraints, near-degenerate states, and state-specific optimization.

\begin{figure*}[htb]
    \centering
    \includegraphics[width=0.75\textwidth]{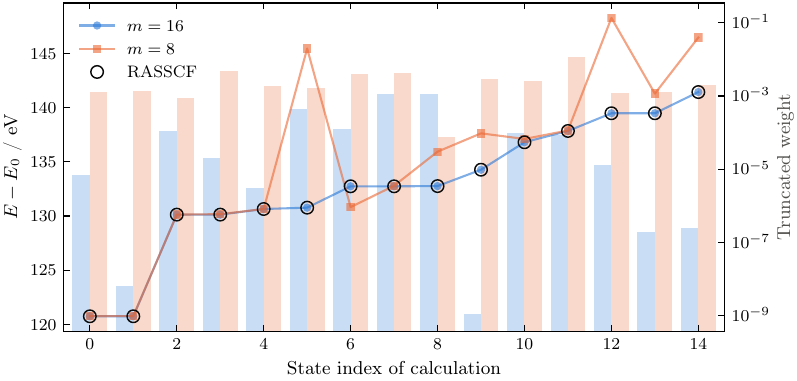}
    \caption{
    The energy of core-excited states with respect to the ground state, $E-E_0$ (left axis, points), in order of appearance in the CVS-ORTHO-DMRG procedure for bond dimensions $m=16$ (blue circles) and $m=8$ (orange squares). 
    For comparison, the RASSCF reference is given as black open circles. 
    Bars show the truncated weights summed over all sweeps: right axis, logarithmic scale, blue $m=16$ and orange $m=8$.}
    \label{fig:ortho_build_up}
\end{figure*}

The simulation of core-level spectra typically requires the calculation of numerous closely spaced electronic states, pushing excited-state DMRG methods toward their practical computational limits. 
The additional numerical difficulties associated with orthogonality-constrained optimization further complicate these calculations.
A typical behavior is illustrated here on an artificially constructed simple case: 
Fig.~\ref{fig:ortho_build_up} displays the energies of the lowest core-excited states, successively obtained in ORTHO-DMRG, for \Be\ with two intentionally low bond dimensions, $m=8$ and 16. 
As a reference, we plot the RASSCF energies, which are free from problems with orthogonally constrained optimization.
The truncated weights are also reported to illustrate the fact that $m=8$ is insufficient to produce reliable results.
For $m=16$, the successive states have increasing energy exactly following the RASSCF reference. 
However, for $m=8$ the ORTHO-DMRG produces a non-monotonic sequence, where most of the states beyond the fourth one are spurious.

In the example discussed above, the bond dimension was deliberately chosen to be sufficiently small to expose these numerical difficulties. However, in our CVS-DMRG studies of more complex systems, which will be reported elsewhere, we have encountered similar problems at substantially larger bond dimensions, with their severity sometimes increasing as $m$ was increased.
These observations suggest the existence of a particularly problematic intermediate regime, in which the bond dimension is already large enough to make the optimization computationally demanding, yet remains insufficient to adequately represent the target states, resulting in substantial truncation errors. In this regime, increasing $m$ does not necessarily improve numerical stability and may even exacerbate convergence difficulties. Once the bond dimension approaches or exceeds the value required for convergence, a further increase generally improves the accuracy and reliability of the results.
Consequently, establishing convergence with respect to $m$ is particularly critical for core-excited states.

Finally, we note that the attempts to employ the shift-and-invert algorithm in our study, and thus avoid using ORTHO-CVS-DMRG, did not lead to satisfactory convergence, presumably due to closely spaced or degenerate states.

\subsection{\texorpdfstring{XAS spectra via \ac{CVS}-\ac{DMRG}}{XAS spectra via CVS-DMRG}}

\subsubsection{\texorpdfstring{$K$-edge of \Be}{K-edge of BeH2}}

\begin{figure*}[htb]
\centering
\includegraphics[width=0.9\textwidth]{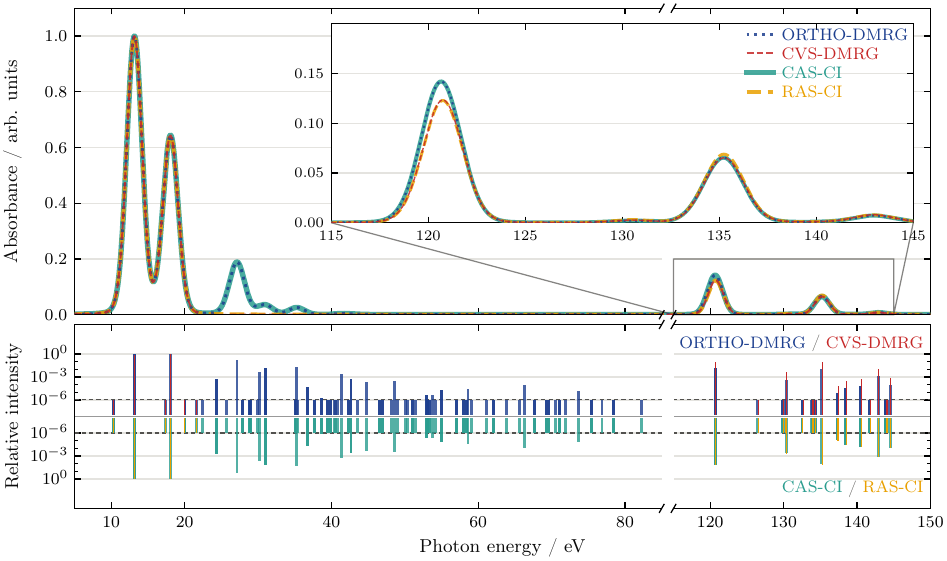}
\caption{
Electronic excitation spectrum of \Be\ across the valence and K-edge energy region, comparing different methods. Note that CVS-DMRG and RAS-CI have fewer valence states. (Upper panel) The broadened spectra using Voigt-profile with Gaussian $\sigma=1$\,eV and Lorentzian 
$\gamma=0.05$\,eV. 
The inset zooms into the K-edge region. 
(Lower panel) Respective stick spectra.
The dashed horizontal lines separate dark and bright states.
}
\label{fig:Be_spectrum}
\end{figure*}

The \ac{CVS}-\ac{DMRG} framework is first validated and illustrated on \Be, whose 
$K$-edge spectrum offers a transparent benchmark owing to the small active space so that complete active space methods, such as \ac{CAS}-\ac{CI} and ORTHO-DMRG, are feasible.
As shown in Fig.~\ref{fig:Be_spectrum}, the full eigenspectrum of the \ac{CAS} Hamiltonian contains a large number of valence-excited states below the core-excited manifold, underscoring the practical necessity of the \ac{CVS} approximation: without it, one would  require to compute and store all lower-lying states before reaching the $K$-edge region, an effort that is both computationally prohibitive and formally unnecessary for the spectroscopic target. The \ac{CVS}-constrained problem
directly accesses the relevant $1s$-excited manifold, dramatically reducing the number of states that must be explicitly resolved. This simplification 
allows us to reproduce the targeted spectral features in  both position and in amplitude as shown in the inset of Fig.~\ref{fig:Be_spectrum}. 
{One sees that the ``exact'' methods CAS-CI and ORTHO-DMRG agree very well with each other. The RAS-CI and CVS-DMRG, which neglect the core-valence correlation, show slight differences from ``exact'' results in energy and intensity for the peak at 121\,eV. Importantly, the results of RAS-CI and CVS-DMRG coincide. Thus, the accuracy of the CVS approximation is very good, while leading to a substantial decrease in computational costs.}

However, such a small system as \Be\ can be calculated in full complexity using the CAS-CI method, and for it all the benefits of CVS-DMRG are not apparent.
The benefits of CVS-DMRG become prominent in cases where a large active space prohibits even the use of core-projected HEXS-RASSCF.

\subsubsection{\texorpdfstring{$L_{2,3}$-, and $M_{3}$-edges of \Fe}{L2,3-, and M3-edges of [Fe(H2O)6]2+}}

\begin{figure*}[htb]
    \centering
    \includegraphics[width=0.9\textwidth]{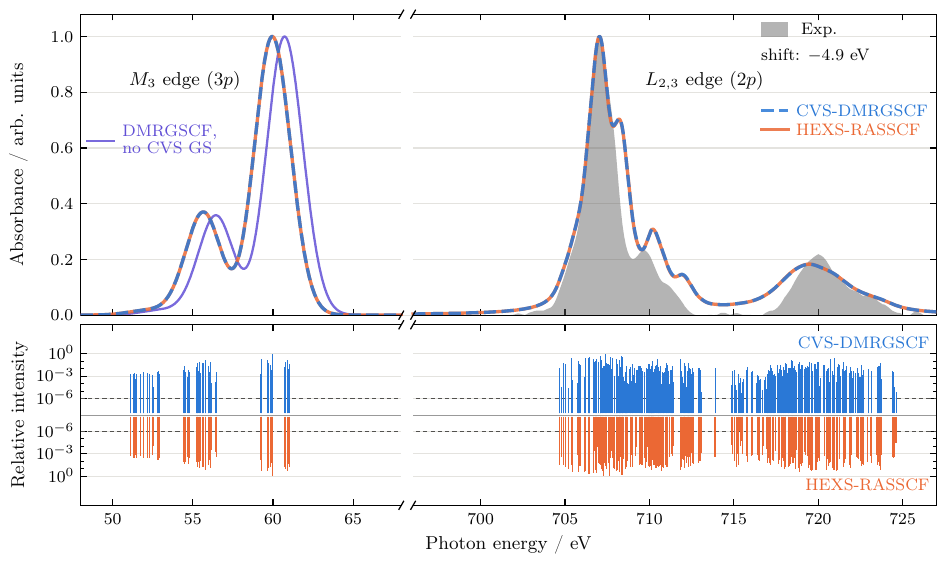}
      \caption{
       Iron $M_3$- (left) and $L_{2,3}$-edge (right) absorption spectra of \Fe:
       (Upper panel) Broadened spectra, each normalized to its own maximum.
       The $M_3$ edge is computed from spin-free quintet states without \ac{SOC} and broadened with a Gaussian profile with $\sigma=1$\,eV; it is plotted without any energy shift.
       The $L_{2,3}$ edge includes \ac{SOC} (triplet and quintet states) and is broadened with $\sigma=0.1$\,eV and $\gamma=0.4$\,eV for $L_3$ and $1.1$\,eV for $L_2$; all $L_{2,3}$ curves are shifted by a common $-4.9$\,eV to fit to the experimetal maximum.
       Besides CVS-DMRGSCF (blue, dashed) and HEXS-RASSCF (orange), the $M_3$ edge shows DMRGSCF without the \ac{CVS} approximation for the ground (valence) states (purple).
       The experimental $L_{2,3}$ spectrum (iPFY)~\cite{atakNatureChemicalBond2013} is shown in gray, scaled to the calculated maximum. 
       (Lower panel) Stick spectra of CVS-DMRGSCF (up) and HEXS-RASSCF (down) on a logarithmic scale of the relative intensity; sticks with intensity below $10^{-6}$ are drawn at the dashed line.
        }
       \label{fig:Fe_spectrum}
\end{figure*}

For spectroscopic applications, accounting for orbital relaxation could be crucial, especially since the presence of a core-hole notably influences the orbitals. 
Therefore, in this section, the performance of the \ac{CVS}-\ac{DMRG}\ac{SCF} approach is demonstrated for the $L_{2,3}$ and  $M_{3}$ edges of \Fe, a prototypical transition-metal complex with a rich X-ray spectrum.
We have chosen these two edges because the $2p^5$ states belong to a deep core, while $3p^5$ states are shallow-core or even deep-valence; comparison of both allows to assess the accuracy of \ac{CVS} approximation for different energetic orbital separations.

Figure~\ref{fig:Fe_spectrum} shows \ac{XAS} for both edges computed with HEXS-RASSCF and CVS-DMRG.
It also compares the calculated spectra with the experiment for the $L_{2,3}$-edge.~\cite{atakNatureChemicalBond2013} 
For the $L$-edge, the CVS results coincide with those of RASSCF and demonstrate good agreement with experiments upon a global shift of $-4.9$\,eV.
The same applies to the $M_3$ edge, where only quintets were computed, and no \ac{SOC} was used; note that there is no published experimental spectrum for this case.
It is remarkable that the spectra and global shift of both methods coincide only when using the CVS Hamiltonian to compute both the valence and core-excited states with $n_{2p}=6$ and 5, respectively.
An example, when the CVS Hamiltonian is used only on the core-excited states, is also given in Fig.~\ref{fig:Fe_spectrum} for the $M$-edge.

For the case of \Fe, we give some methodological remarks on the computational protocol.
In particular, we have studied the role of the double-shell $4d$ orbitals in the active space, which were found to be important to retrieve the radial $d$-orbital correlation for the valence states.~\cite{botchValenceCorrelation21981,Anderson_CPL_1992} 
This effect can be particularly important for $L$-edge spectra, where the number of $d$ electrons changes.
The results can be found in {Sec.~S5} of the \supp.
Removing the $4d$ orbitals from the active space, AS(12e,8o), leaves the shape of the $L_{2,3}$ spectrum essentially unchanged: once each spectrum is aligned to the experiment, the two broadened spectra and the intensities in stick spectra ({Fig.~\ref{fig:Fe_spectrum}}) nearly coincide. 
The double-shell orbitals affect mainly the absolute position of the edge, with the global alignment shifts differing by about 1\,eV ($-4.91$\,eV with $4d$ versus $-5.88$\,eV without them).
Remarkably, the energy lowering of each state due to the inclusion of $4d$ orbitals linearly correlates with their occupation for each state; see {Fig.~S6} of the \supp. 
In general, in the $2p^5 3d^7$ core-excited states, the $4d$ orbitals are populated more than in the valence $2p^6 3d^6$ states, 
owing to the stronger dynamic correlation of the additional $d$ electron. Therefore, they are stabilized more (by about 2.5--3\,eV versus 1.7--2.1\,eV for valence states). 
Thus, the double-shell effect manifests as an almost state-independent stabilization within each core manifold,
which is absorbed by the global energy shift applied to compare with experiment.

Additional stabilization occurs due to orbital relaxation. In 
Figs.~{S7} and S8 of the \supp, 
we compare the results computed using the fixed ground-state orbital basis for all state manifolds, denoted as HEXS-RAS-CI, and the relaxed basis for each manifold.
For the deep $2p$ core hole,  
relaxing the orbitals lowers the core-excited states by 2.24--2.45\,eV, 
which is not entirely uniform---the $\sim$0.2\,eV variation among the 30 roots slightly reshapes the level structure.
At the $M$ edge, ground-state orbitals give nearly the same results as the relaxed ones.
Therefore, orbital relaxation cannot be neglected for deep core-hole states.

\subsection{QIT characterization of CVS}
\label{sec:QIT}

\begin{figure}[htb]
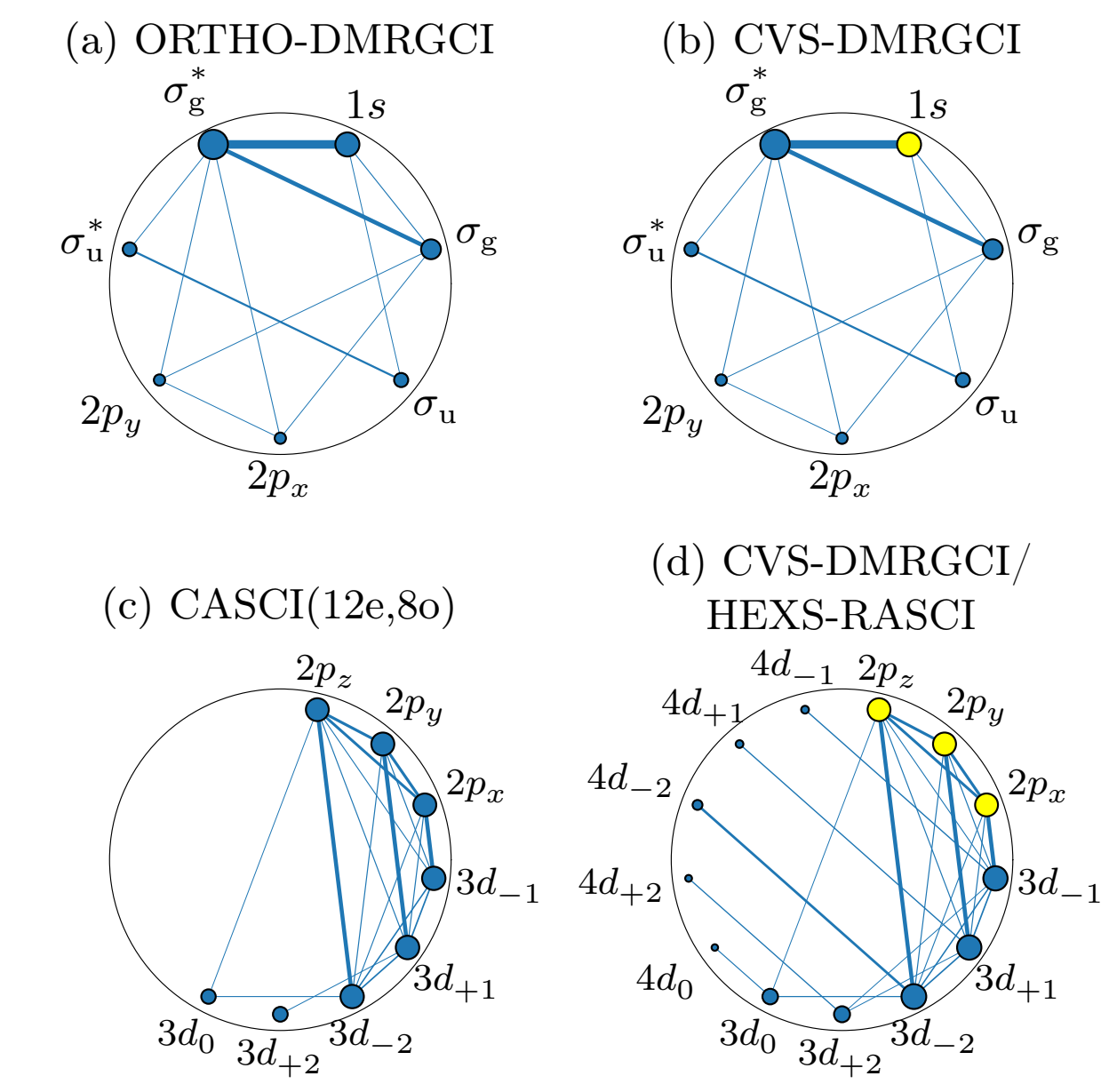

    \centering
    \Entanglement
    \caption{
       Single-orbital entanglement entropies (size of the vertices) and mutual information (thickness of the edges).
       (a,b) The lowest \Be\ $1s$-excited eigenvector obtained by ORTHO-\ac{DMRG} and \ac{CVS}-\ac{DMRG}CI.
       (c,d) Lowest \Fe\ $2p$-excited state computed with CAS-CI without HEXS projection (in an active space excluding the $4d$ orbitals) and with \ac{CVS}-\ac{DMRG}CI. All calculations use the same set of orbitals, optimized for this state. The HEXS-RASCI results coincide with those of \ac{CVS}-\ac{DMRG}CI.
       Yellow circles denote the core orbital used in the \ac{CVS} approach. 
    }
    \label{fig:clocks}
\end{figure}

To establish a rigorous theoretical understanding of how the \ac{CVS} approximation influences the many-body wavefunction, we analyze the core-excited eigenstates through the lens of quantum information theory.
Fig.~\ref{fig:clocks} compares the single-orbital entanglement entropies 
\[s_i^{(1)}=-\text{Tr}(\rho_i\ln\rho_i)\] 
and mutual information 
\[I_{ij}=\tfrac{1}{2}\left[s_i^{(1)}+s_j^{(1)}-s_{ij}^{(2)}\right]\,,\]
of the lowest \Be\ $1s$-excited and \Fe\ $2p$-excited states. 
Here the two-orbital entropy is 
\[s_{ij}^{(2)}=-\text{Tr}(\rho_{ij}\ln\rho_{ij})\,,\] 
and $\rho_i$ and $\rho_{ij}$ refer to one- and two-particle reduced density matrices.

The left column (panels (a) and (c)) shows the \ac{QIT} quantities obtained by the reference method without applying the CVS.
For \Be, it is the fully correlated ORTHO-\ac{DMRG} and the projection-free (non-HEXS) CAS-CI for \Fe.
In the latter case, only the use of a smaller active space (12e,8o) was feasible.
The right column represents the \ac{CVS}-\ac{DMRG}CI 
for \Be\ and \Fe.  
Note also that the results of HEXS-RASSCF (not shown) visually coincide with those of CVS-DMRGSCF in panel (d).

The two sets of $s_i^{(1)}$ and $I_{ij}$ diagnostics for reference and CVS methods are nearly identical. For instance, for the $1s$ orbital of \Be, the single-orbital entropy differs by less than $1.2\%$ 
and the total pairwise correlation 
by less than $1.0\%$; 
the numerical values can be found in Table~S2 in \supp.
The results of CVS-DMRG (and HEXS-RAS-CI) for \Fe\ cannot be directly compared to unconstrained ORTHO-DMRG because of the prohibitively large number of states to be computed.
Instead, we compare to CAS-CI in a smaller active space (12e,8o), because the larger one (12e,13o) was again not feasible at this level of theory.
Despite missing $4d$ orbitals in panel (c), one sees that the entanglement pattern is visually unchanged for the rest of the active space.

This quantitative agreement demonstrates that \ac{CVS}-\ac{DMRG} produces eigenstates whose orbital entanglement structure is very similar to that of the unconstrained references.  
Therefore, the CVS approximation does not quench essential correlation between the core and the valence orbitals. 
To assess the quality of the CVS approximation, a more informative quantity to examine is the expectation value and variance of the core number operator.

\begin{figure}[b]
    \centering
    \includegraphics[width=\columnwidth]{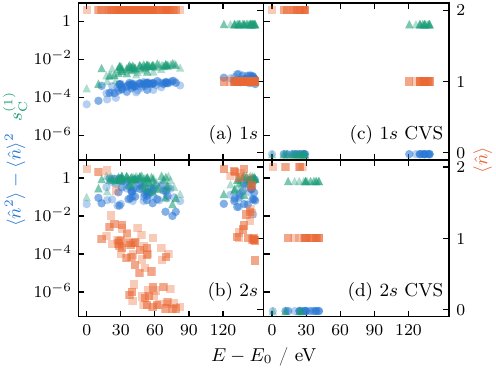}
    \caption{ 
    Single-orbital entropies, $s^{(1)}_C$ green triangles, 
    of the $1s$, (a) and (c), and $2s$, (b) and (d), orbitals, the expectation value, $\braket{\hat n}$ orange squares, and variance, $\braket{\hat n^2}-\braket{\hat n}^2$ blue circles, of the core-electron number operator for the \Be\ system in the electronic states ordered by excitation energy, $E-E_0$.
    The left column corresponds to the same CAS-CI calculation with 135 states, but the operators in rows correspond to $1s$- and $2s$-orbital quantities. The right column shows the values for the CVS-DMRG calculations where $1s$ or $2s$ orbitals are defined as core. 
    }
    \label{fig:Be_numbers}
\end{figure}

{The respective analysis, shown in Fig.~\ref{fig:Be_numbers} for the \Be\  $(1s)^{-1}$ and $(2s)^{-1}$ states, fully illustrates the effect of the \ac{CVS} approximation and indicates cases where it works better or worse. 
In  panel (a), the expectation value and the variance of the $1s$ number operator are computed on the first 135 eigenstates of the full \ac{CAS}-\ac{CI} Hamiltonian, of which only the highest 30 are core-excited states. 
One can see a drop of $\braket{\hat n}$ by one and an energy jump when the core-excitation threshold is reached. 
This is also marked by the increase of the core-orbital entropy $s^{(1)}_\text{C}$ from {nearly zero} for the double occupation to about $\ln 2$ upon appearance of the core-hole {due to spin entanglement of the core spin}.
At the same time, the variance $\braket{\hat n^2}-\braket{\hat n}^2$ stays relatively low across the whole spectrum. 
In panel (c), one can see that \ac{CVS}  imposes constraints and strengthens these two properties, as the eigenvectors have all zero variance and integer population.

In panel (b), one can see the results of the same ORTHO-DMRG calculation for the 135 states, but now the deep valence $2s\sigma$ orbital is in focus. 
The expectation value $\braket{\hat n_{2s}}$ is not an integer anymore and goes from 2 to 0, where the clusters with $(2s\sigma)^{-1}$ (valence and core) and $(2s\sigma)^{-2}$ (valence) can be seen. The variance, in turn, increases by about two orders of magnitude. 
However, if one defines $2s\sigma$ as core in CVS, panel (d),  
one again enforces the zero variance and integer population.
Although for the upper panels ($1s$ core) CVS just ``purified'' the situation and allowed us to address core-excited states directly, for the lower panels ($2s\sigma$ core) 
CVS 
represents a crude approximation.
There is also {an apparent shift in energy} when one compares panels (c) and (d) in the range 0--50\,eV, emphasizing the very approximate nature of CVS for the $2s$ orbital.

\begin{figure}[tb]
    \centering
    \includegraphics[width=\columnwidth]{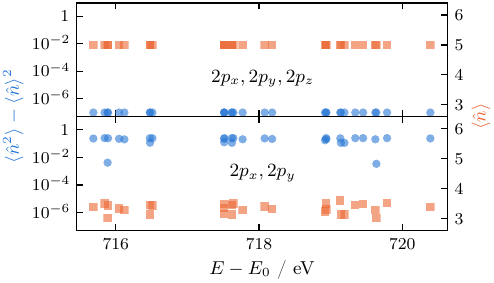}
    \caption{Expectation values and variance of the number operator for the $2p$-hole states of the \Fe\ system for the total $2p$ number operator (upper panel) and  the number operator that includes only $2p_x$ and $2p_y$ orbitals (bottom panel): expectation value, $\braket{\hat n}$ orange squares (right axis), and variance, $\braket{\hat n^2}-\braket{\hat n}^2$ blue circles (left axis). 
     }
    \label{fig:Fe_numbers}
\end{figure}

Further, Fig.~\ref{fig:Fe_numbers} illustrates a somewhat obvious fact which we would still like to emphasize: for the cases when multiple orbitals are set to be core in CVS, the occupations of individual orbitals are not constrained due to strong exchange interaction between core electrons. 
For instance, for \Fe\ $(2p)^5$ configurations the $\hat n_{2p_x}+\hat n_{2p_y}+\hat n_{2p_z}$ is integer and shows no variance, whereas $\hat n_{2p_x}+\hat n_{2p_y}$ demonstrates both dispersion in occupation numbers and large variance. 
This fact has to be kept in mind when dealing with systems having equivalent or nearly equivalent atoms in a molecule, and thus multiple (nearly-)degenerate core orbitals, suggesting that all such orbitals should be included. Moreover, in general, orbitals  cannot be excluded from the CVS core due to symmetry, as symmetry selection rules, although being relevant for transition strengths, do not apply to exchange couplings, and such an exclusion can compromise the accuracy of the electronic structure.

Finally, we note that CVS-DMRGCI/SCF and HEXS-RAS-CI/SCF produce nearly identical results both in view of spectra and QIT quantities.
Although the core-valence separation is performed following two different routes for these methods, i.e., neglect of certain Hamiltonian terms in CVS and projecting out valence configurations in HEXS, these methods can be considered practically identical (for converged bond dimension $m$).
However, CVS-DMRG provides access to substantially larger active spaces than HEXS-RAS methods by exploiting a compact wavefunction representation achieved through controlled truncation of the Schmidt spectrum.

\subsection{Perturbative correction to the CVS energy}

\begin{figure}[b]
    \centering
    \includegraphics[width=0.95\columnwidth]{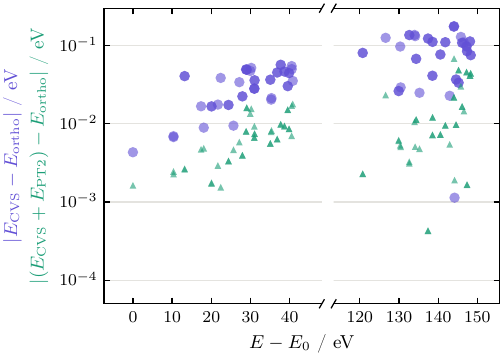}
    \caption{
    The second-order perturbation corrections to the $1s$ CVS energies in \Be.
    The ORTHO-DMRG ($E_\text{ortho}$) is used as a reference: absolute errors of the CVS energies (purple circles) and of the CVS+PT2 energies (green triangles).
    {States are ordered according to their CVS/CVS+PT2 excitation energy $E-E_0$, respectively.}
    }
    \label{fig:Be_PT2}
\end{figure}

\begin{figure*}[htb]
    \centering
   \includegraphics[width=0.9\textwidth]{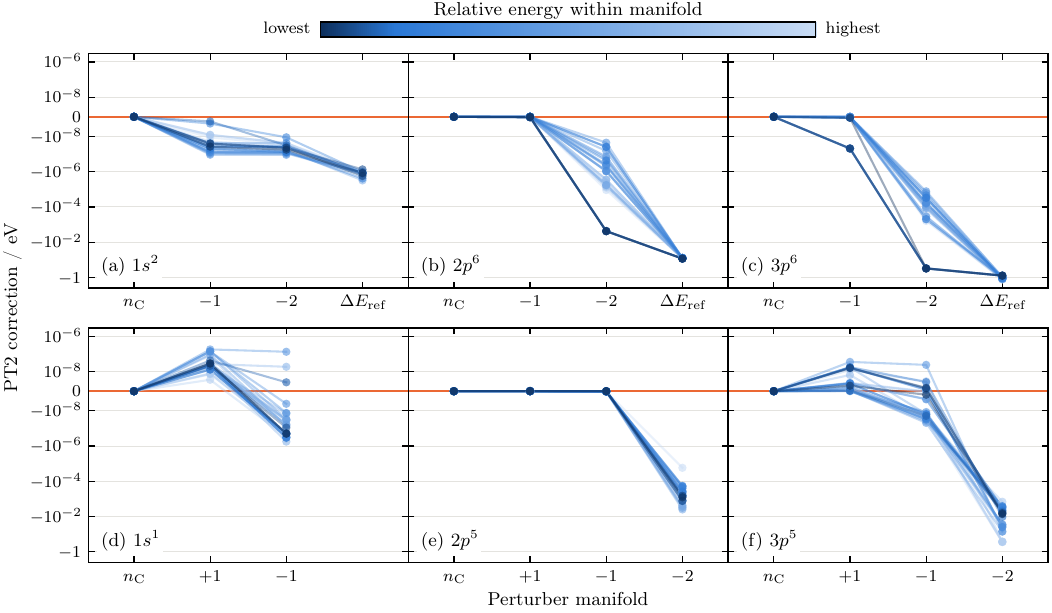}
   \caption{
   Second-order perturbation correction to the CVS energy for the \Fe\ system for the lowest 30 quintet states in each manifold: 
   (a) lowest valence states with CVS applied to $1s$ orbital with $n_\text{C}=2$; (b) the same valence states but for $2p$ and $n_\text{C}=6$; (c) the same valence states for $3p$ and $n_\text{C}=6$; the lower panels (d), (e), and (f) represent different manifolds of $1s$ (K-edge), $2p$ (L-edge), and $3p$ (M-edge) core-hole states with $n_\text{C}$ being 1, 5, and 5, respectively.
   The color of the 30 lines corresponds to the relative energetic position of the state within a manifold with respect to the lowest state in this manifold.
   The \ac{PT2} correction is given with respect to the CVS reference energy of each state represented by the {red} solid line.
   The horizontal axis denotes the contributions of all states $J$, see Eq.~\eqref{eq:PT2}, to the PT2 energy with the given electron occupation difference from the original $n_\text{C}$. 
   The corrections are applied successively: first, bare CVS, then $+1$ on top of it, $-1$, and, finally, $-2$ manifolds.
   The {$\Delta E_\text{ref}$} corresponds to the difference between the CVS-DMRG reference energy and the ORTHO-DMRG result.
   }
    \label{fig:Fe_PT2}
\end{figure*}

The \ac{CVS} approximation introduces a systematic energy error by restricting the Hamiltonian. 
To recover a significant portion of this error and to provide a quantitative estimate of its magnitude, we apply the second-order perturbative correction presented in Sec.~\ref{subsec:PT2_theory}. 
Again, the \Be\ is chosen as it allows a comparison with the exact case, while \Fe demonstrates the generality of the method. 

As shown in Fig.~\ref{fig:Be_PT2}, for $1s$ edge of \Be, the CVS error is 
of the order of 0.1\,eV.
Adding the \ac{PT2} correction makes the ground and valence-excited state energies lie closer to those of ORTHO-DMRG. For example, the error is reduced by approximately one order of magnitude mainly uniformly across the $K$-edge spectrum of \Be. 
{Importantly, since we apply the correction on the energies but not on the \ac{CVS}-optimized \ac{MPS} itself, the spectral intensities -- governed by the \ac{CVS} transition moments -- remain at the \ac{CVS} level of accuracy.}

The same procedure is applied to various core levels of \Fe and the results are shown in Fig.~\ref{fig:Fe_PT2}, where the individual contributions from each core population manifold are resolved separately. 
The upper row, panels (a), (b), and (c), displays the same set of 30 quintet valence states where different orbitals are set to be core in the CVS treatment; the respective orbitals remain, however, fully occupied.
The lower row, panels (d), (e), and (f), corresponds to the states with a core hole in $1s$, $2p$, and $3p$ orbitals, respectively.
The contributions with the respective $n_\text{C}$
are taken as a reference, and the energy corrections due to different $\mathcal{M}(n_\text{C}',n_\text{V}')$-manifolds, Eq.~\eqref{eq:PT2}, with $n_\text{C}\pm1$ and $n_\text{C}-2$  are given along the x-axis; $n_\text{C}+2$ is forbidden by the Pauli principle.
{The corrections are applied successively: first, bare CVS, then +1 on top of it, then -1, and, finally, -2 manifolds.}

For the valence states, panels (a)--(c), the correction due to core-excited states in general lowers the energy; panel (b) is an exception where $n_\text{C}-1$ states lead to almost no change.
The account for the PT2 correction, thus, improves the agreement with the ORTHO-DMRG reference, denoted as $\Delta E_\text{ref}$ in the figure.
This improvement is especially pronounced for the lower-lying states {(darker colors)}.
The higher-lying states {(lighter colors)} would presumably need more than 30 perturber states in each manifold to get closer to the ORTHO-DMRG reference.
Note also that for deep $1s$ core-hole states the overall discrepancy from ORTHO-DMRG and the total magnitude of the PT2 correction are of order of $10^{-8}-10^{-6}$\,eV.
For less deep $2p$ core-hole states, it is several orders of magnitude higher, and finally, for the deep valence $3p$ holes, both the $\Delta E_\text{ref}$ and the PT2 correction reach about 1\,eV.

For core states, panels (d)--(f), adding the valence manifold in general raises the energy, and accounting for the $-1$ and $-2$ manifolds does not lead to the homogeneous energy corrections as in the case of valence states.
Overall, the magnitude of the total correction also increases in the row $1s$, $2p$, $3p$, indicating that the bare CVS approximation is fairly good for $K$- and $L$-edges and much less accurate for the $M$-edge.
Unfortunately, the direct ORTHO-DMRG comparison in this case is not feasible.
The effect of the \ac{PT2} correction on the $M_3$-edge spectrum of \Fe\ is shown in  
{Fig.~S11} in \supp.

In conclusion, CVS-\ac{PT2} correction can serve two primary purposes: it improves absolute excitation energies, and it provides an internally consistent, reference-free diagnostic of the \ac{CVS} error magnitude that can be evaluated without access to an unconstrained calculation. 

\FloatBarrier

\section{Conclusions and outlook}
\label{sec:conclusions}

In this work, we have presented the formulation and implementation of the \ac{CVS} approximation within the \ac{DMRG} framework, establishing \ac{CVS}-\ac{DMRG} as a practical and efficient method for the ab initio simulation of core-level X-ray spectra in strongly correlated systems.
Two symmetry implementations were suggested. The \uf\ scheme
was shown to be overly restrictive when more than one orbital is set to be core.
The \ut\ scheme
resolves this deficiency and produces spin-pure solutions; 
it is recommended as the main \ac{CVS} implementation.
Finally, the generalization to a multi-core \ac{CVS} Hamiltonian, in which each of $N$ orbital subspaces conserves its occupation independently, was formulated to enable the simultaneous targeting of excitations from multiple core shells within a single \ac{DMRG} calculation.

The accuracy of the \ac{CVS} approximation was assessed on \Be\ by direct comparison with \ac{CAS}-\ac{CI} and ORTHO-\ac{DMRG}, confirming that \ac{CVS}-\ac{DMRG} faithfully reproduces both excitation energies and spectral intensities across the $K$-edge manifold. The method was then applied to the substantially larger active space of \Fe, where HEXS-RASSCF serves as a benchmark, demonstrating that \ac{CVS}-\ac{DMRG} gives results in excellent agreement with it at the $L$- and $M$-edges while remaining applicable to active spaces that are beyond the reach of \ac{RAS}-based approaches. 

A quantum-information analysis of the \ac{CVS} eigenstates provided a rigorous characterization of the approximation. The central finding is that 
the \ac{CVS} constraint enforces integer core occupations and zero number operator variance, while leaving core--valence orbital entanglement nearly unaltered. This both explains why \ac{CVS} performs reliably when applied to deep core orbitals and provides a transparent diagnostic for identifying orbitals for which the approximation would introduce an uncontrolled error.

Finally, a second-order perturbative correction to the \ac{CVS} energies was derived and applied. The correction couples the \ac{CVS} eigenstates to configurations in adjacent core-electron manifolds through the residual $\hat{H}' = \hat{H} - \hat{H}^{\text{CVS}}$, and reduces the \ac{CVS} energy error.
Crucially, the correction is reference-free: it requires no unconstrained calculation and therefore provides an internally consistent estimate of the \ac{CVS} error for any system and any edge.

Several directions remain open for future work. The combination of \ac{CVS}-\ac{DMRG} with internally contracted perturbation theories such as \ac{NEV} \ac{PT2}~\cite{Guo_JCTC_2016,Freitag_JCTC_2017} or \ac{CAS}\ac{PT2}~\cite{Kurashige_JCP_2011,Wouters_JCP_2016} would provide a systematic route to dynamic correlation beyond the perturbative correction introduced here. Finally, the extension to multiple core orbitals paves the way for investigating simultaneous core excitations and multi-site core-ionized states,~\cite{Pelimanni2024} which have so far remained largely inaccessible to conventional multi-configurational approaches.

\section*{Acknowledgements}

G.B. and S.I.B. gratefully acknowledge the financial support from Deutsche Forschungsgemeinschaft Grants No. 521855965 and 521855798.  M.R. gratefully acknowledges support by the Swiss National Science Foundation through grant no. 200021\_219616.

\section*{Data availability statement}

Data for this article, {including the calculation output files with full echo of the input files, HDF5 files with wave functions needed to perform these calculations, and the data for figures in this publication}, are available at Zenodo at https://doi.org/10.5281/zenodo.24019316.

\section*{Author Declarations}

The authors have no conflicts to disclose.

\bibliography{ZZ_final_bib}

@article{Widmark_TCA_1990,
  title = {Density Matrix Averaged Atomic Natural Orbital ({{ANO}}) Basis Sets for Correlated Molecular Wave Functions},
  author = {Widmark, Per-Olof and Malmqvist, Per {\AA}ke and Roos, Bj{\"o}rn O.},
  year = 1990,
  journal = {Theor. Chim. Acta},
  volume = {77},
  pages = {291--306},
  doi = {10.1007/BF01120130},
  langid = {english}
}

@article{Hess_PRA_1986,
  title = {Relativistic Electronic-Structure Calculations Employing a Two-Component No-Pair Formalism with External-Field Projection Operators},
  author = {Hess, Bernd A.},
  year = 1986,
  journal = {Phys. Rev. A},
  volume = {33},
  number = {6},
  pages = {3742},
  doi = {10.1103/PhysRevA.33.3742},
  urldate = {2014-06-14}
}

@article{Roos_JPCA_2005,
  title = {New {{Relativistic ANO Basis Sets}} for {{Transition Metal Atoms}}},
  author = {Roos, Bj{\"o}rn O. and Lindh, Roland and Malmqvist, Per-{\AA}ke and Veryazov, Valera and Widmark, Per-Olof},
  year = 2005,
  month = jul,
  journal = {J. Phys. Chem. A},
  volume = {109},
  number = {29},
  pages = {6575--6579},
  issn = {1089-5639, 1520-5215},
  doi = {10.1021/jp0581126},
  urldate = {2014-05-22}
}

@article{Douglas_AP_1974,
  title = {Quantum Electrodynamical Corrections to the Fine Structure of Helium},
  author = {Douglas, Marvin and Kroll, Norman M.},
  year = 1974,
  journal = {Ann. Phys.},
  volume = {82},
  number = {1},
  pages = {89--155},
  doi = {10.1016/0003-4916(74)90333-9},
  urldate = {2014-06-14}
}

@article{Pelimanni2024,
  author = {Pelimanni, Eetu and
            Fouda, Adam E. A. and
            Ho, Phay J. and
            Baumann, Thomas M. and
            Bokarev, Sergey I. and
            De Fanis, Alberto and
            Dold, Simon and
            Grell, Gilbert and
            Ismail, Iyas and
            Koulentianos, Dimitrios and
            Mazza, Tommaso and
            Meyer, Michael and
            Piancastelli, Maria-Novella and
            P{\"u}ttner, Ralph and
            Rivas, Daniel E. and
            Senfftleben, Bj{\"o}rn and
            Simon, Marc and
            Young, Linda and
            Doumy, Gilles},
  title   = {Observation of Molecular Resonant Double-Core
             Excitation Driven by Intense {X}-ray Pulses},
  journal = {Communications Physics},
  volume  = {7},
  number  = {1},
  pages   = {341},
  year    = {2024},
  doi     = {10.1038/s42005-024-01804-5}
}

@article{Wouters_JCP_2016,
    author = {Wouters, Sebastian and Van Speybroeck, Veronique and Van Neck, Dimitri},
    title = {DMRG-CASPT2 study of the longitudinal static second hyperpolarizability of all-trans polyenes},
    journal = {The Journal of Chemical Physics},
    volume = {145},
    number = {5},
    pages = {054120},
    year = {2016},
    month = {08},
    issn = {0021-9606},
    doi = {10.1063/1.4959817},
    url = {https://doi.org/10.1063/1.4959817},
}

@article{Kurashige_JCP_2011,
    author = {Kurashige, Yuki and Yanai, Takeshi},
    title = {Second-order perturbation theory with a density matrix renormalization group self-consistent field reference function: Theory and application to the study of chromium dimer},
    journal = {The Journal of Chemical Physics},
    volume = {135},
    number = {9},
    pages = {094104},
    year = {2011},
    month = {09},
    issn = {0021-9606},
    doi = {10.1063/1.3629454},
    url = {https://doi.org/10.1063/1.3629454},
}

@article{Freitag_JCTC_2017,
    author = {Freitag, Leon and Knecht, Stefan and Angeli, Celestino and Reiher, Markus},
    title = {Multireference Perturbation Theory with Cholesky Decomposition
for the Density Matrix Renormalization Group},
    journal = {Journal of Chemical Theory and Computation},
    volume = {13},
    number = {2},
    pages = {451-459},
    year = {2017},
    month = {02},
    issn = {1549-9618},
    doi = {10.1021/acs.jctc.6b00778},
    url = {https://doi.org/10.1021/acs.jctc.6b00778},
}

@article{Guo_JCTC_2016,
    author = {Guo, Sheng and Watson, Mark A. and Hu, Weifeng and Sun, Qiming and Chan, Garnet Kin-Lic},
    title = {N‑Electron Valence State Perturbation
Theory Based on a Density Matrix Renormalization Group Reference Function,
with Applications to the Chromium Dimer and a Trimer Model of Poly(p‑Phenylenevinylene)},
    journal = {Journal of Chemical Theory and Computation},
    volume = {12},
    number = {4},
    pages = {1583-1591},
    year = {2016},
    month = {04},
    issn = {1549-9618},
    doi = {10.1021/acs.jctc.5b01225},
    url = {https://doi.org/10.1021/acs.jctc.5b01225},
}

@article{Anderson_CPL_1992,
title = {Excitation energies in the nickel atom studied with the complete active space SCF method and second-order perturbation theory},
journal = {Chemical Physics Letters},
volume = {191},
number = {6},
pages = {507-514},
year = {1992},
issn = {0009-2614},
doi = {https://doi.org/10.1016/0009-2614(92)85581-T},
url = {https://www.sciencedirect.com/science/article/pii/000926149285581T},
author = {Kerstin Andersson and Björn O. Roos}
}

@article{Freitag_JCTC_2021,
    author = {Freitag, Leon and Baiardi, Alberto and Knecht, Stefan and Gonz{\'a}lez, Leticia},
    title = {Simplified State Interaction for Matrix Product State
Wave Functions},
    journal = {Journal of Chemical Theory and Computation},
    volume = {17},
    number = {12},
    pages = {7477-7485},
    year = {2021},
    month = {12},
    issn = {1549-9618},
    doi = {10.1021/acs.jctc.1c00674},
    url = {https://doi.org/10.1021/acs.jctc.1c00674},
}

@article{Knecht_JCTC_2016,
    author = {Knecht, Stefan and Keller, Sebastian and Autschbach, Jochen and Reiher, Markus},
    title = {A Nonorthogonal State-Interaction
Approach for Matrix
Product State Wave Functions},
    journal = {Journal of Chemical Theory and Computation},
    volume = {12},
    number = {12},
    pages = {5881-5894},
    year = {2016},
    month = {12},
    issn = {1549-9618},
    doi = {10.1021/acs.jctc.6b00889},
    url = {https://doi.org/10.1021/acs.jctc.6b00889},
}

@article{Malmqvist_CPL_2002,
  title = {The Restricted Active Space ({{RAS}}) State Interaction Approach with Spin--Orbit Coupling},
  author = {Malmqvist, Per {\AA}ke and Roos, Bj{\"o}rn O. and Schimmelpfennig, Bernd},
  year = 2002,
  month = may,
  journal = {Chemical Physics Letters},
  volume = {357},
  number = {3-4},
  pages = {230--240},
  issn = {00092614},
  doi = {10.1016/S0009-2614(02)00498-0},
  urldate = {2024-07-26},
  copyright = {https://www.elsevier.com/tdm/userlicense/1.0/},
  langid = {english}
}

@misc{PythonScripts,
  author       = {Bassi, Giacomo},
  title        = {Python tools for assessment of {CVS-DMRG} accuracy},
  year         = {2026},
  howpublished = {\url{https://github.com/Giacomo-Bassi/pyQIT.git}},
}

@article{Catarina_EPJB_2023,
  title = {Density-Matrix Renormalization Group: A Pedagogical Introduction},
  shorttitle = {Density-Matrix Renormalization Group},
  author = {Catarina, G. and Murta, Bruno},
  year = 2023,
  month = aug,
  journal = {Eur. Phys. J. B},
  volume = {96},
  number = {8},
  pages = {111},
  issn = {1434-6036},
  doi = {10.1140/epjb/s10051-023-00575-2},
  urldate = {2026-10-17},
  langid = {english}
}

@article{Jahnke_CR_2020,
    author = {Jahnke, Till and Hergenhahn, Uwe and Winter, Bernd and Dörner, Reinhard and Frühling, Ulrike and Demekhin, Philipp V. and Gokhberg, Kirill and Cederbaum, Lorenz S. and Ehresmann, Arno and Knie, André and Dreuw, Andreas},
    title = {Interatomic and Intermolecular Coulombic Decay},
    journal = {Chemical Reviews},
    volume = {120},
    number = {20},
    pages = {11295-11369},
    year = {2020},
    month = {10},
    issn = {0009-2665},
    doi = {10.1021/acs.chemrev.0c00106},
    url = {https://doi.org/10.1021/acs.chemrev.0c00106},
}

@article{Wodynski_JCTC_2018,
    author = {Wodyński, Artur and Kaupp, Martin},
    title = {Noncollinear Two-Component Quasirelativistic Description
of Spin Interactions in Exchange-Coupled Systems. Mapping Generalized
Broken-Symmetry States to Effective Spin Hamiltonians},
    journal = {Journal of Chemical Theory and Computation},
    volume = {14},
    number = {3},
    pages = {1267-1276},
    year = {2018},
    month = {03},
    issn = {1549-9618},
    doi = {10.1021/acs.jctc.7b01067},
    url = {https://doi.org/10.1021/acs.jctc.7b01067},
}

@article{Safari_JCTC_2025,
    author = {Safari, Arta A. and Bogdanov, Nikolay A.},
    title = {Effective Hamiltonians
from Spin-Adapted Configuration
Interaction},
    journal = {Journal of Chemical Theory and Computation},
    volume = {21},
    number = {2},
    pages = {539-548},
    year = {2025},
    month = {01},
    issn = {1549-9618},
    doi = {10.1021/acs.jctc.4c01380},
    url = {https://doi.org/10.1021/acs.jctc.4c01380},
}

@article{CheMPS2,
title = {CheMPS2: A free open-source spin-adapted implementation of the density matrix renormalization group for ab initio quantum chemistry},
journal = {Computer Physics Communications},
volume = {185},
number = {6},
pages = {1501-1514},
year = {2014},
issn = {0010-4655},
doi = {https://doi.org/10.1016/j.cpc.2014.01.019},
url = {https://www.sciencedirect.com/science/article/pii/S0010465514000496},
author = {Sebastian Wouters and Ward Poelmans and Paul W. Ayers and Dimitri {Van Neck}}
}

@article{Block2,
    author = {Zhai, Huanchen and Larsson, Henrik R. and Lee, Seunghoon and Cui, Zhi-Hao and Zhu, Tianyu and Sun, Chong and Peng, Linqing and Peng, Ruojing and Liao, Ke and Tölle, Johannes and Yang, Junjie and Li, Shuoxue and Chan, Garnet Kin-Lic},
    title = {Block2: A comprehensive open source framework to develop and apply state-of-the-art DMRG algorithms in electronic structure and beyond},
    journal = {The Journal of Chemical Physics},
    volume = {159},
    number = {23},
    pages = {234801},
    year = {2023},
    month = {12},
    issn = {0021-9606},
    doi = {10.1063/5.0180424},
    url = {https://doi.org/10.1063/5.0180424},
}

@article{Tenorio_JCTC_2022,
    author = {Tenorio, Bruno Nunes Cabral and Voß, Torben Arne and Bokarev, Sergey I. and Decleva, Piero and Coriani, Sonia},
    title = {Multireference Approach to Normal and Resonant Auger
Spectra Based on the One-Center Approximation},
    journal = {Journal of Chemical Theory and Computation},
    volume = {18},
    number = {7},
    pages = {4387-4407},
    year = {2022},
    month = {07},
    issn = {1549-9618},
    doi = {10.1021/acs.jctc.2c00252},
    url = {https://doi.org/10.1021/acs.jctc.2c00252},
}

@article{Grell_JCP_2020,
    author = {Grell, Gilbert and Bokarev, Sergey I.},
    title = {Multi-reference protocol for (auto)ionization spectra: Application to molecules},
    journal = {The Journal of Chemical Physics},
    volume = {152},
    number = {7},
    pages = {074108},
    year = {2020},
    month = {02},
    issn = {0021-9606},
    doi = {10.1063/1.5142251},
    url = {https://doi.org/10.1063/1.5142251},
}

@article{Grell_PRA_2019,
  title = {Multireference quantum chemistry protocol for simulating autoionization spectra: Test of ionization continuum models for the neon atom},
  author = {Grell, Gilbert and K{\"u}hn, Oliver and Bokarev, Sergey I.},
  journal = {Phys. Rev. A},
  volume = {100},
  issue = {4},
  pages = {042512},
  numpages = {14},
  year = {2019},
  month = {Oct},
  publisher = {American Physical Society},
  doi = {10.1103/PhysRevA.100.042512},
  url = {https://link.aps.org/doi/10.1103/PhysRevA.100.042512}
}

@article{Norman_CR_2018,
    author = {Norman, Patrick and Dreuw, Andreas},
    title = {Simulating X‑ray Spectroscopies and Calculating
Core-Excited States of Molecules},
    journal = {Chemical Reviews},
    volume = {118},
    number = {15},
    pages = {7208-7248},
    year = {2018},
    month = {08},
    issn = {0009-2665},
    doi = {10.1021/acs.chemrev.8b00156},
    url = {https://doi.org/10.1021/acs.chemrev.8b00156},
}

@article{deGroot_JESRP_2021,
title = {2p x-ray absorption spectroscopy of 3d transition metal systems},
journal = {Journal of Electron Spectroscopy and Related Phenomena},
volume = {249},
pages = {147061},
year = {2021},
issn = {0368-2048},
doi = {https://doi.org/10.1016/j.elspec.2021.147061},
url = {https://www.sciencedirect.com/science/article/pii/S0368204821000190},
author = {Frank M.F. {de Groot} and Hebatalla Elnaggar and Federica Frati and Ru-pan Wang and Mario U. Delgado-Jaime and Michel {van Veenendaal} and Javier Fernandez-Rodriguez and Maurits W. Haverkort and Robert J. Green and Gerrit {van der Laan} and Yaroslav Kvashnin and Atsushi Hariki and Hidekazu Ikeno and Harry Ramanantoanina and Claude Daul and Bernard Delley and Michael Odelius and Marcus Lundberg and Oliver Kuhn and Sergey I. Bokarev and Eric Shirley and John Vinson and Keith Gilmore and Mauro Stener and Giovanna Fronzoni and Piero Decleva and Peter Kruger and Marius Retegan and Yves Joly and Christian Vorwerk and Claudia Draxl and John Rehr and Arata Tanaka},
}

@article{Lischka_CR_2018,
    author = {Lischka, Hans and Nachtigallová, Dana and Aquino, Adélia J.
A. and Szalay, Péter G. and Plasser, Felix and Machado, Francisco B. C. and Barbatti, Mario},
    title = {Multireference Approaches for Excited States of Molecules},
    journal = {Chemical Reviews},
    volume = {118},
    number = {15},
    pages = {7293-7361},
    year = {2018},
    month = {08},
    issn = {0009-2665},
    doi = {10.1021/acs.chemrev.8b00244},
    url = {https://doi.org/10.1021/acs.chemrev.8b00244},
}

@article{Khedkar_PCCP_2021,
    author = {Khedkar, Abhishek and Roemelt, Michael},
    title = {Modern multireference methods and their application in transition metal chemistry},
    journal = {Physical Chemistry Chemical Physics},
    volume = {23},
    number = {32},
    pages = {17097-17112},
    year = {2021},
    month = {08},
    issn = {1463-9076},
    doi = {10.1039/d1cp02640b},
    url = {https://doi.org/10.1039/d1cp02640b},
}

@article{Park_CR_2020,
    author = {Park, Jae Woo and Al-Saadon, Rachael and MacLeod, Matthew K. and Shiozaki, Toru and Vlaisavljevich, Bess},
    title = {Multireference Electron Correlation Methods: Journeys
along Potential Energy Surfaces},
    journal = {Chemical Reviews},
    volume = {120},
    number = {13},
    pages = {5878-5909},
    year = {2020},
    month = {07},
    issn = {0009-2665},
    doi = {10.1021/acs.chemrev.9b00496},
    url = {https://doi.org/10.1021/acs.chemrev.9b00496},
}

@article{FEAST2009,
  title = {Density-matrix-based algorithm for solving eigenvalue problems},
  author = {Polizzi, Eric},
  journal = {Phys. Rev. B},
  volume = {79},
  issue = {11},
  pages = {115112},
  numpages = {6},
  year = {2009},
  month = {Mar},
  publisher = {American Physical Society},
  doi = {10.1103/PhysRevB.79.115112},
  url = {https://link.aps.org/doi/10.1103/PhysRevB.79.115112}
}

@article{kellerSelectionActiveSpaces2015,
  title = {Selection of Active Spaces for Multiconfigurational Wavefunctions},
  author = {Keller, Sebastian and Boguslawski, Katharina and Janowski, Tomasz and Reiher, Markus and Pulay, Peter},
  date = {2015-06-23},
  year = 2015,
  journal = {The Journal of Chemical Physics},
  shortjournal = {J. Chem. Phys.},
  volume = {142},
  number = {24},
  pages = {244104},
  issn = {0021-9606},
  doi = {10.1063/1.4922352},
  url = {https://doi.org/10.1063/1.4922352},
}

@article{bokarevTheoreticalXraySpectroscopy2020,
  title = {Theoretical {{X}}-ray Spectroscopy of Transition Metal Compounds},
  author = {Bokarev, Sergey I. and K{\"u}hn, Oliver},
  year = 2020,
  month = jan,
  journal = {WIREs Computational Molecular Science},
  volume = {10},
  number = {1},
  pages = {e1433},
  issn = {1759-0876, 1759-0884},
  doi = {10.1002/wcms.1433},
  langid = {english}
}

@article{aliverti-piuriWhatCanQuantum2024,
  title = {What Can Quantum Information Theory Offer to Quantum Chemistry?},
  author = {{Aliverti-Piuri}, Damiano and Chatterjee, Kaustav and Ding, Lexin and Liao, Ke and Liebert, Julia and Schilling, Christian},
  year = 2024,
  journal = {Faraday Discussions},
  volume = {254},
  number = {0},
  pages = {76--106},
  publisher = {Royal Society of Chemistry},
  doi = {10.1039/D4FD00059E},
  langid = {english}
}

@article{White_PRB_1993,
  title = {Density-Matrix Algorithms for Quantum Renormalization Groups},
  author = {White, Steven R.},
  date = {1993-10-01},
  year = 1993,
  journal = {Physical Review B},
  shortjournal = {Phys. Rev. B},
  volume = {48},
  number = {14},
  pages = {10345--10356},
  publisher = {American Physical Society},
  doi = {10.1103/PhysRevB.48.10345},
  url = {https://link.aps.org/doi/10.1103/PhysRevB.48.10345},
}

@article{Schollwock_AoP_2011a,
  title = {The Density-Matrix Renormalization Group in the Age of Matrix Product States},
  author = {Schollwöck, Ulrich},
  date = {2011-01-01},
  year = 2011,
  journal = {Annals of Physics},
  shortjournal = {Annals of Physics},
  series = {January 2011 {{Special Issue}}},
  volume = {326},
  number = {1},
  pages = {96--192},
  issn = {0003-4916},
  doi = {10.1016/j.aop.2010.09.012},
  url = {https://www.sciencedirect.com/science/article/pii/S0003491610001752},
}

@article{Baiardi_JCP_2020b,
  title = {The Density Matrix Renormalization Group in Chemistry and Molecular Physics: {{Recent}} Developments and New Challenges},
  shorttitle = {The Density Matrix Renormalization Group in Chemistry and Molecular Physics},
  author = {Baiardi, Alberto and Reiher, Markus},
  date = {2020-01-23},
  year = 2020,
  journal = {The Journal of Chemical Physics},
  shortjournal = {J. Chem. Phys.},
  volume = {152},
  number = {4},
  pages = {040903},
  issn = {0021-9606},
  doi = {10.1063/1.5129672},
  url = {https://doi.org/10.1063/1.5129672},
}

@article{chanDensityMatrixRenormalization2011,
  title = {The {{Density Matrix Renormalization Group}} in {{Quantum Chemistry}}},
  author = {Chan, Garnet Kin-Lic and Sharma, Sandeep},
  year = 2011,
  month = may,
  journal = {Annual Review of Physical Chemistry},
  volume = {62},
  number = {Volume 62, 2011},
  pages = {465--481},
  publisher = {Annual Reviews},
  issn = {0066-426X, 1545-1593},
  doi = {10.1146/annurev-physchem-032210-103338},
  langid = {english}
}

@article{Booth_JCP_2009,
  title = {Fermion {{Monte Carlo}} without Fixed Nodes: {{A}} Game of Life, Death, and Annihilation in {{Slater}} Determinant Space},
  shorttitle = {Fermion {{Monte Carlo}} without Fixed Nodes},
  author = {Booth, George H. and Thom, Alex J. W. and Alavi, Ali},
  date = {2009-08-04},
  year = 2009,
  journal = {The Journal of Chemical Physics},
  shortjournal = {J. Chem. Phys.},
  volume = {131},
  number = {5},
  pages = {054106},
  issn = {0021-9606},
  doi = {10.1063/1.3193710},
  url = {https://doi.org/10.1063/1.3193710},
}

@article{LiManni_JCTC_2023b,
  title = {The {{OpenMolcas Web}}: {{A Community-Driven Approach}} to {{Advancing Computational Chemistry}}},
  shorttitle = {The {{OpenMolcas Web}}},
  author = {Li Manni, Giovanni and Fdez. Galv{\'a}n, Ignacio and Alavi, Ali and Aleotti, Flavia and Aquilante, Francesco and Autschbach, Jochen and Avagliano, Davide and Baiardi, Alberto and Bao, Jie J. and Battaglia, Stefano and Birnoschi, Letitia and {Blanco-Gonz{\'a}lez}, Alejandro and Bokarev, Sergey I. and Broer, Ria and Cacciari, Roberto and Calio, Paul B. and Carlson, Rebecca K. and Carvalho Couto, Rafael and Cerd{\'a}n, Luis and Chibotaru, Liviu F. and Chilton, Nicholas F. and Church, Jonathan Richard and Conti, Irene and Coriani, Sonia and {Cu{\'e}llar-Zuquin}, Juliana and Daoud, Razan E. and Dattani, Nike and Decleva, Piero and {de Graaf}, Coen and Delcey, Micka{\"e}l G. and De Vico, Luca and Dobrautz, Werner and Dong, Sijia S. and Feng, Rulin and Ferr{\'e}, Nicolas and Filatov(Gulak), Michael and Gagliardi, Laura and Garavelli, Marco and Gonz{\'a}lez, Leticia and Guan, Yafu and Guo, Meiyuan and Hennefarth, Matthew R. and Hermes, Matthew R. and Hoyer, Chad E. and {Huix-Rotllant}, Miquel and Jaiswal, Vishal Kumar and Kaiser, Andy and Kaliakin, Danil S. and Khamesian, Marjan and King, Daniel S. and Kochetov, Vladislav and Kro{\'s}nicki, Marek and Kumaar, Arpit Arun and Larsson, Ernst D. and Lehtola, Susi and Lepetit, Marie-Bernadette and Lischka, Hans and L{\'o}pez R{\'i}os, Pablo and Lundberg, Marcus and Ma, Dongxia and Mai, Sebastian and Marquetand, Philipp and Merritt, Isabella C. D. and Montorsi, Francesco and M{\"o}rchen, Maximilian and Nenov, Artur and Nguyen, Vu Ha Anh and Nishimoto, Yoshio and Oakley, Meagan S. and Olivucci, Massimo and Oppel, Markus and Padula, Daniele and Pandharkar, Riddhish and Phung, Quan Manh and Plasser, Felix and Raggi, Gerardo and Rebolini, Elisa and Reiher, Markus and Rivalta, Ivan and {Roca-Sanju{\'a}n}, Daniel and Romig, Thies and Safari, Arta Anushirwan and {S{\'a}nchez-Mansilla}, Aitor and Sand, Andrew M. and Schapiro, Igor and Scott, Thais R. and {Segarra-Mart{\'i}}, Javier and Segatta, Francesco and Sergentu, Dumitru-Claudiu and Sharma, Prachi and Shepard, Ron and Shu, Yinan and Staab, Jakob K. and Straatsma, Tjerk P. and S{\o}rensen, Lasse Kragh and Tenorio, Bruno Nunes Cabral and Truhlar, Donald G. and Ungur, Liviu and Vacher, Morgane and Veryazov, Valera and Vo{\ss}, Torben Arne and Weser, Oskar and Wu, Dihua and Yang, Xuchun and Yarkony, David and Zhou, Chen and Zobel, J. Patrick and Lindh, Roland},
  year = 2023,
  month = oct,
  journal = {Journal of Chemical Theory and Computation},
  volume = {19},
  number = {20},
  pages = {6933--6991},
  publisher = {American Chemical Society},
  issn = {1549-9618},
  doi = {10.1021/acs.jctc.3c00182},
}

@article{Szenes_JPCA_2025,
  title = {{{QCMaquis}} 4.0: {{Multipurpose Electronic}}, {{Vibrational}}, and {{Vibronic Structure}} and {{Dynamics Calculations}} with the {{Density Matrix Renormalization Group}}},
  shorttitle = {{{QCMaquis}} 4.0},
  author = {Szenes, Kalman and Glaser, Nina and Erakovic, Mihael and Barandun, Valentin and M{\"o}rchen, Maximilian and Feldmann, Robin and Battaglia, Stefano and Baiardi, Alberto and Reiher, Markus},
  year = 2025,
  month = aug,
  journal = {The Journal of Physical Chemistry A},
  volume = {129},
  number = {32},
  pages = {7549--7574},
  publisher = {American Chemical Society},
  issn = {1089-5639},
  doi = {10.1021/acs.jpca.5c02970},
}

@article{butscherModificationDavidsonsMethod1976,
  title = {Modification of {{Davidson}}'s Method for the Calculation of Eigenvalues and Eigenvectors of Large Real-Symmetric Matrices: “Root Homing Procedure”},
  shorttitle = {Modification of {{Davidson}}'s Method for the Calculation of Eigenvalues and Eigenvectors of Large Real-Symmetric Matrices},
  author = {Butscher, W. and Kammer, W.E.},
  year = 1976,
  date = {1976-03},
  journal = {Journal of Computational Physics},
  shortjournal = {Journal of Computational Physics},
  volume = {20},
  number = {3},
  pages = {313--325},
  issn = {00219991},
  doi = {10.1016/0021-9991(76)90084-X},
  url = {https://linkinghub.elsevier.com/retrieve/pii/002199917690084X},
  langid = {english}
}

@article{rakhubaCalculatingVibrationalSpectra2016,
  title = {Calculating Vibrational Spectra of Molecules Using Tensor Train Decomposition},
  author = {Rakhuba, Maxim and Oseledets, Ivan},
  year = 2016,
  month = sep,
  journal = {The Journal of Chemical Physics},
  volume = {145},
  number = {12},
  pages = {124101},
  issn = {0021-9606},
  doi = {10.1063/1.4962420},
}

@book{yousefsaadNumericalMethodsLarge,
  title = {Numerical Methods for Large Eigenvalue Problems},
  author = {Saad, Yousef},
  year = {2011},
  publisher = {{Society for Industrial and Applied Mathematics}},
  doi = {10.1137/1.9781611970739},
  url = {https://epubs.siam.org/doi/abs/10.1137/1.9781611970739}
}

@article{jiaHarmonicRefinedHarmonic2015,
  title = {Harmonic and Refined Harmonic Shift-Invert Residual {{Arnoldi}} and {{Jacobi}}--{{Davidson}} Methods for Interior Eigenvalue Problems},
  author = {Jia, Zhongxiao and Li, Cen},
  year = 2015,
  month = jul,
  journal = {Journal of Computational and Applied Mathematics},
  volume = {282},
  pages = {83--97},
  issn = {0377-0427},
  doi = {10.1016/j.cam.2014.12.043}
}

@article{dorandoTargetedExcitedState2007a,
  title = {Targeted Excited State Algorithms},
  author = {Dorando, Jonathan J. and Hachmann, Johannes and Chan, Garnet Kin-Lic},
  date = {2007-08-28},
  year = 2007,
  journal = {The Journal of Chemical Physics},
  shortjournal = {J. Chem. Phys.},
  volume = {127},
  number = {8},
  pages = {084109},
  issn = {0021-9606},
  doi = {10.1063/1.2768360},
  url = {https://doi.org/10.1063/1.2768360},
}

@article{Olsen_JCP_1988b,
  title = {Determinant Based Configuration Interaction Algorithms for Complete and Restricted Configuration Interaction Spaces},
  author = {Olsen, Jeppe and Roos, Björn O. and Jo/rgensen, Poul and Jensen, Hans Jo/rgen Aa.},
  date = {1988-08-15},
  year = 1988,
  journal = {The Journal of Chemical Physics},
  shortjournal = {J. Chem. Phys.},
  volume = {89},
  number = {4},
  pages = {2185--2192},
  issn = {0021-9606},
  doi = {10.1063/1.455063},
  url = {https://doi.org/10.1063/1.455063},
}

@article{Malmqvist_JPC_1990a,
  title = {The Restricted Active Space Self-Consistent-Field Method, Implemented with a Split Graph Unitary Group Approach},
  author = {Malmqvist, Per Aake. and Rendell, {\relax Alistair}. and Roos, Bjoern O.},
  year = 1990,
  month = jul,
  journal = {The Journal of Physical Chemistry},
  volume = {94},
  number = {14},
  pages = {5477--5482},
  publisher = {American Chemical Society},
  issn = {0022-3654},
  doi = {10.1021/j100377a011},
}

@article{Ma_JCP_2011a,
  title = {The Generalized Active Space Concept in Multiconfigurational Self-Consistent Field Methods},
  author = {Ma, Dongxia and Li Manni, Giovanni and Gagliardi, Laura},
  year = 2011,
  month = jul,
  journal = {The Journal of Chemical Physics},
  volume = {135},
  number = {4},
  pages = {044128},
  issn = {0021-9606},
  doi = {10.1063/1.3611401},
}

@article{Nakano_CPL_2000a,
  title = {A Quasi-Complete Active Space Self-Consistent Field Method},
  author = {Nakano, Haruyuki and Hirao, Kimihiko},
  year = 2000,
  month = jan,
  journal = {Chemical Physics Letters},
  volume = {317},
  number = {1},
  pages = {90--96},
  issn = {0009-2614},
  doi = {10.1016/S0009-2614(99)01364-0},
}

@article{Ivanic_JCP_2003a,
  title = {Direct Configuration Interaction and Multiconfigurational Self-Consistent-Field Method for Multiple Active Spaces with Variable Occupations. {{I}}. {{Method}}},
  author = {Ivanic, Joseph},
  year = 2003,
  month = nov,
  journal = {The Journal of Chemical Physics},
  volume = {119},
  number = {18},
  pages = {9364--9376},
  issn = {0021-9606},
  doi = {10.1063/1.1615954},
}

@article{delceyEfficientCalculationsLarge2019,
  title = {Efficient Calculations of a Large Number of Highly Excited States for Multiconfigurational Wavefunctions},
  author = {Delcey, Mickael G. and Sørensen, Lasse Kragh and Vacher, Morgane and Couto, Rafael C. and Lundberg, Marcus},
  year = {2019},
  journal = {Journal of Computational Chemistry},
  volume = {40},
  number = {19},
  pages = {1789--1799},
  issn = {1096-987X},
  doi = {10.1002/jcc.25832},
  url = {https://onlinelibrary.wiley.com/doi/abs/10.1002/jcc.25832},
  langid = {english}
}

@article{Cederbaum_PRA_1980b,
  title = {Many-Body Theory of Core Holes},
  author = {Cederbaum, L. S. and Domcke, W. and Schirmer, J.},
  year = 1980,
  month = jul,
  journal = {Physical Review A},
  volume = {22},
  number = {1},
  pages = {206--222},
  publisher = {American Physical Society},
  doi = {10.1103/PhysRevA.22.206},
}

@article{atakNatureChemicalBond2013,
  title = {},
  author = {Atak, Kaan and Bokarev, Sergey I. and Gotz, Malte and Golnak, Ronny and Lange, Kathrin M. and Engel, Nicholas and Dantz, Marcus and Suljoti, Edlira and Kühn, Oliver and Aziz, Emad F.},
  date = {2013-10-17},
  journal = {J. Phys. Chem. B},
  shortjournal = {J. Phys. Chem. B},
  volume = {117},
  number = {41},
  pages = {12613--12618},
  publisher = {American Chemical Society},
  issn = {1520-6106},
  doi = {10.1021/jp408212u},
  url = {https://doi.org/10.1021/jp408212u},
  urldate = {2026-04-23}
}

@article{botchValenceCorrelation21981,
  title = {Valence Correlation in the {\emph{s}} 2 {\emph{d}} {\emph{n}} , {\emph{s}} {\emph{d}} {\emph{n}} +1, and {\emph{d}} {\emph{n}} +2 States of the First-Row Transition Metal Atoms},
  author = {Botch, Beatrice H. and Dunning, {\relax Thom}. H. and Harrison, James F.},
  year = 1981,
  month = oct,
  journal = {The Journal of Chemical Physics},
  volume = {75},
  number = {7},
  pages = {3466--3476},
  issn = {0021-9606, 1089-7690},
  doi = {10.1063/1.442456},
  urldate = {2026-09-01},
  langid = {english}
}

@article{Barth_PRA_1981b,
  title = {Many-Body Theory of Core-Valence Excitations},
  author = {Barth, A. and Cederbaum, L. S.},
  date = {1981-03-01},
  year = 1981,
  journal = {Physical Review A},
  shortjournal = {Phys. Rev. A},
  volume = {23},
  number = {3},
  pages = {1038--1061},
  publisher = {American Physical Society},
  doi = {10.1103/PhysRevA.23.1038},
  url = {https://link.aps.org/doi/10.1103/PhysRevA.23.1038}, 
}

@article{Coriani_PRA_2012,
  title = {Coupled-Cluster Response Theory for near-Edge x-Ray-Absorption Fine Structure of Atoms and Molecules},
  author = {Coriani, Sonia and Christiansen, Ove and Fransson, Thomas and Norman, Patrick},
  date = {2012-02-08},
  year = 2012,
  journal = {Physical Review A},
  shortjournal = {Phys. Rev. A},
  volume = {85},
  number = {2},
  pages = {022507},
  publisher = {American Physical Society},
  doi = {10.1103/PhysRevA.85.022507},
  url = {https://link.aps.org/doi/10.1103/PhysRevA.85.022507},  
}

@article{Schirmer_PRA_1982b,
  title = {Beyond the Random-Phase Approximation: {{A}} New Approximation Scheme for the Polarization Propagator},
  shorttitle = {Beyond the Random-Phase Approximation},
  author = {Schirmer, Jochen},
  date = {1982-11-01},
  year = 1982,
  journal = {Physical Review A},
  shortjournal = {Phys. Rev. A},
  volume = {26},
  number = {5},
  pages = {2395--2416},
  publisher = {American Physical Society},
  doi = {10.1103/PhysRevA.26.2395},
  url = {https://link.aps.org/doi/10.1103/PhysRevA.26.2395},
}

@article{Mandal_JCP_2024,
  title = {A New Parameterization of the {{DFT}}/{{CIS}} Method with Applications to Core-Level Spectroscopy},
  author = {Mandal, Aniket and Berquist, Eric J. and Herbert, John M.},
  year = 2024,
  month = jul,
  journal = {The Journal of Chemical Physics},
  volume = {161},
  number = {4},
  pages = {044114},
  issn = {0021-9606},
  doi = {10.1063/5.0220535},
}

@article{Sadybekov_JCP_2017,
  title = {Coupled-Cluster Based Approach for Core-Level States in Condensed Phase: {{Theory}} and Application to Different Protonated Forms of Aqueous Glycine},
  shorttitle = {Coupled-Cluster Based Approach for Core-Level States in Condensed Phase},
  author = {Sadybekov, Arman and Krylov, Anna I.},
  year = 2017,
  month = jul,
  journal = {The Journal of Chemical Physics},
  volume = {147},
  number = {1},
  pages = {014107},
  issn = {0021-9606},
  doi = {10.1063/1.4990564},
}

@article{Norman_CR_2018b,
  title = {Simulating {{X-ray Spectroscopies}} and {{Calculating Core-Excited States}} of {{Molecules}}},
  author = {Norman, Patrick and Dreuw, Andreas},
  year = 2018,
  month = aug,
  journal = {Chemical Reviews},
  volume = {118},
  number = {15},
  pages = {7208--7248},
  publisher = {American Chemical Society},
  issn = {0009-2665},
  doi = {10.1021/acs.chemrev.8b00156},
}

@article{Wormit_MP_2014,
  title = {Investigating Excited Electronic States Using the Algebraic Diagrammatic Construction ({{ADC}}) Approach of the Polarisation Propagator},
  author = {Wormit, Michael and Rehn, Dirk R. and Harbach, Philipp H.P. and Wenzel, Jan and Krauter, Caroline M. and Epifanovsky, Evgeny and Dreuw, Andreas},
  year = 2014,
  month = mar,
  journal = {Molecular Physics},
  volume = {112},
  number = {5-6},
  pages = {774--784},
  publisher = {Taylor \& Francis},
  issn = {0026-8976},
  doi = {10.1080/00268976.2013.859313}
}

@article{kellerEfficientMatrixProduct2015,
  title = {An Efficient Matrix Product Operator Representation of the Quantum Chemical {{Hamiltonian}}},
  author = {Keller, Sebastian and Dolfi, Michele and Troyer, Matthias and Reiher, Markus},
  date = {2015-12-28},
  year = 2015,
  journal = {The Journal of Chemical Physics},
  volume = {143},
  number = {24},
  pages = {244118},
  issn = {0021-9606, 1089-7690},
  doi = {10.1063/1.4939000},
  url = {https://pubs.aip.org/jcp/article/143/24/244118/967039/An-efficient-matrix-product-operator},
  langid = {english}
}

@article{baiardiOptimizationHighlyExcited2019,
  title = {Optimization of Highly Excited Matrix Product States with an Application to Vibrational Spectroscopy},
  author = {Baiardi, Alberto and Stein, Christopher J. and Barone, Vincenzo and Reiher, Markus},
  date = {2019-03-07},
  year = 2019,
  journal = {The Journal of Chemical Physics},
  volume = {150},
  number = {9},
  pages = {094113},
  issn = {0021-9606, 1089-7690},
  doi = {10.1063/1.5068747},
  url = {https://pubs.aip.org/jcp/article/150/9/094113/76063/Optimization-of-highly-excited-matrix-product},
  langid = {english}
}

@article{baiardiExcitedStateDMRGMade2022,
  title = {Excited-{{State DMRG Made Simple}} with {{FEAST}}},
  author = {Baiardi, Alberto and Kelemen, Anna Kl{\'a}ra and Reiher, Markus},
  year = 2022,
  month = jan,
  journal = {Journal of Chemical Theory and Computation},
  volume = {18},
  number = {1},
  pages = {415--430},
  publisher = {American Chemical Society},
  issn = {1549-9618},
  doi = {10.1021/acs.jctc.1c00984},
}

@book{trygvehelgakeMolecularElectronicStructureTheory2013,
  title = {Molecular {{Electronic-Structure Theory}}},
  author = {{Trygve Helgake} and {Poul Jorgensen} and {Jeppe Olsen}},
  year = {2013},
  publisher = {Wiley}
}

@article{McCulloch_JSM_2007,
  title = {From Density-Matrix Renormalization Group to Matrix Product States},
  author = {McCulloch, Ian P},
  year = 2007,
  month = oct,
  journal = {Journal of Statistical Mechanics: Theory and Experiment},
  volume = {2007},
  number = {10},
  pages = {P10014},
  issn = {1742-5468},
  doi = {10.1088/1742-5468/2007/10/P10014},
  langid = {english}
}

@article{kellerSpinadaptedMatrixProduct2016,
  title = {Spin-Adapted {{Matrix Product States}} and {{Operators}}},
  author = {Keller, Sebastian and Reiher, Markus},
  year = 2016,
  month = apr,
  journal = {The Journal of Chemical Physics},
  volume = {144},
  number = {13},
  pages = {134101},
  issn = {0021-9606, 1089-7690},
  doi = {10.1063/1.4944921},
  archiveprefix = {arXiv}
}

@article{Coriani_JCP_2015b,
  title = {Communication: {{X-ray}} Absorption Spectra and Core-Ionization Potentials within a Core-Valence Separated Coupled Cluster Framework},
  shorttitle = {Communication},
  author = {Coriani, Sonia and Koch, Henrik},
  year = 2015,
  month = nov,
  journal = {The Journal of Chemical Physics},
  volume = {143},
  number = {18},
  pages = {181103},
  issn = {0021-9606},
  doi = {10.1063/1.4935712},
}

@article{boguslawskiOrbitalEntanglementQuantum2015,
  title = {Orbital Entanglement in Quantum Chemistry},
  author = {Boguslawski, Katharina and Tecmer, Paweł},
  year = {2015},
  journal = {International Journal of Quantum Chemistry},
  volume = {115},
  number = {19},
  pages = {1289--1295},
  issn = {1097-461X},
  doi = {10.1002/qua.24832},
  url = {https://onlinelibrary.wiley.com/doi/abs/10.1002/qua.24832},
  langid = {english}
}

@article{rankineProgressTheoryXray2021,
  title = {Progress in the {{Theory}} of {{X-ray Spectroscopy}}: {{From Quantum Chemistry}} to {{Machine Learning}} and {{Ultrafast Dynamics}}},
  shorttitle = {Progress in the {{Theory}} of {{X-ray Spectroscopy}}},
  author = {Rankine, C. D. and Penfold, T. J.},
  date = {2021-05-27},
  year = 2021,
  journal = {The Journal of Physical Chemistry A},
  shortjournal = {J. Phys. Chem. A},
  volume = {125},
  number = {20},
  pages = {4276--4293},
  publisher = {American Chemical Society},
  issn = {1089-5639},
  doi = {10.1021/acs.jpca.0c11267},
  url = {https://doi.org/10.1021/acs.jpca.0c11267},
}

@article{hachmannRadicalCharacterAcenes2007,
  title = {The Radical Character of the Acenes: {{A}} Density Matrix Renormalization Group Study},
  shorttitle = {The Radical Character of the Acenes},
  author = {Hachmann, Johannes and Dorando, Jonathan J. and Avil{\'e}s, Michael and Chan, Garnet Kin-Lic},
  year = 2007,
  month = oct,
  journal = {The Journal of Chemical Physics},
  volume = {127},
  number = {13},
  pages = {134309},
  issn = {0021-9606},
  doi = {10.1063/1.2768362},
  urldate = {2026-03-13}
}

@article{aravenaPeriodicTrendsLanthanide2016,
  title = {Periodic {{Trends}} in {{Lanthanide Compounds}} through the {{Eyes}} of {{Multireference}} Ab {{Initio Theory}}},
  author = {Aravena, Daniel and Atanasov, Mihail and Neese, Frank},
  year = 2016,
  month = may,
  journal = {Inorganic Chemistry},
  volume = {55},
  number = {9},
  pages = {4457--4469},
  publisher = {American Chemical Society},
  issn = {0020-1669},
  doi = {10.1021/acs.inorgchem.6b00244},
  urldate = {2026-03-13}
}

@article{baroneAccurateStructureThermodynamics2013,
  title = {Accurate Structure, Thermodynamics, and Spectroscopy of Medium-Sized Radicals by Hybrid Coupled Cluster/Density Functional Theory Approaches: {{The}} Case of Phenyl Radical},
  shorttitle = {Accurate Structure, Thermodynamics, and Spectroscopy of Medium-Sized Radicals by Hybrid Coupled Cluster/Density Functional Theory Approaches},
  author = {Barone, Vincenzo and Biczysko, Malgorzata and Bloino, Julien and Egidi, Franco and Puzzarini, Cristina},
  year = 2013,
  month = jun,
  journal = {The Journal of Chemical Physics},
  volume = {138},
  number = {23},
  pages = {234303},
  issn = {0021-9606},
  doi = {10.1063/1.4810863},
  urldate = {2026-03-13}
}
\end{document}